\documentclass{aa} 

\usepackage{graphicx}
\usepackage{txfonts}
\usepackage{hyperref}
\usepackage{amsmath}
\usepackage{rotating}
\usepackage[usenames,dvipsnames]{color}

\newcommand{\sub}[1]{\ensuremath{_\mathrm{#1}}}

\newcommand{\mbeq}{\overset{!}{=}}

\newcommand{\Rev}[1]{#1}

\DeclareFontFamily{U}{euc}{}
\DeclareFontShape{U}{euc}{m}{n}{<-6>eurm5<6-8>eurm7<8->eurm10}{}%
\DeclareSymbolFont{AMSc}{U}{euc}{m}{n} 
\DeclareMathSymbol{\umu}{\mathord}{AMSc}{"16} 

\defcitealias{paper1}{Paper I}
\defcitealias{mulders10}{M10}

\begin{document}

   \title{The architecture of the LkCa 15 transitional disk\\
   		revealed by high-contrast imaging}

    \titlerunning{\Rev{LkCa 15 disk architecture}}


   \author{C. Thalmann\inst{1,2}
          \and
          G. D. Mulders\inst{2,3}
          \and
          K. Hodapp\inst{4}
          \and
          M. Janson\inst{5}
          \and
          C. A. Grady\inst{6}\fnmsep\inst{7}\fnmsep\inst{8}
          \and
          M. Min\inst{2}
          \and
          M. de Juan Ovelar\inst{2,9}
          \and
          J. Carson\inst{10}
          \and
          T. Brandt\inst{11}
          \and
          M. Bonnefoy\inst{12}
          \and
          M. W. McElwain\inst{7}
          \and
          J. Leisenring\inst{13}
          \and
          C. Dominik\inst{2}
          \and
          T. Henning\inst{12}
          \and
          M. Tamura\inst{14,15}
          }

   \institute{Institute for Astronomy, ETH Zurich, Wolfgang-Pauli-Strasse 27, 8093 Zurich, Switzerland\\
            \email{thalmann@phys.ethz.ch}
         	\and
            Astronomical Institute ``Anton Pannekoek'', University of Amsterdam, 
       		Science Park 904, 1098 XH Amsterdam, The Netherlands
			\and
            Lunar and Planetary Laboratory, The University of Arizona, 1629 E. University Blvd., Tucson, AZ 85721, USA 
            \and
			Institute for Astronomy, University of Hawaii, 2680 Woodlawn Drive, 
			Honolulu, HI 96822, USA
			\and
            Astrophysics Research Center, Queen's University Belfast, Belfast,
            Northern Ireland, UK
            \and
            Eureka Scientific, 2452 Delmer, Suite 100, Oakland CA 96002, USA
			\and
			Exoplanets and Stellar Astrophysics Laboratory, Code 667, Goddard Space 
			Flight Center, Greenbelt, MD 20771, USA
			\and
			Goddard Center for Astrobiology, Goddard Space Flight Center, Greenbelt, 
			MD 20771, USA
            \and
            Leiden Observatory, Leiden University, P.O. Box 9513, 
            2300RA Leiden, The Netherlands
    		\and
			Department of Physics \& Astronomy, College of Charleston, 58 Coming Street, Charleston, SC 29424, USA
            \and
			Astrophysics Department, Institute for Advanced Study, 
			Princeton, NJ, USA
            \and
    		Max Planck Institute for Astronomy, K\"onigstuhl 17, 
            69117 Heidelberg, Germany
            \and
            Department of Astronomy and Steward Observatory, 933 North Cherry Avenue,
            Rm. N204, Tucson, AZ 85721-0065, USA
            \and
            National Astronomical Observatory of Japan, 2-21-1 Osawa,
            Mitaka, Tokyo 181-8588, Japan
            \and
            Department of Astronomical Science, Graduate University for
            Advanced Studies (Sokendai), Tokyo 181-8588, Japan
            }

   \date{Draft version}

 
  \abstract{We present four new epochs of $K\sub{s}$-band images of the young
pre-transitional disk around LkCa~15, and perform extensive forward
modeling to derive the physical parameters of the disk.  We find
indications of strongly anisotropic scattering (Henyey-Greenstein \Rev{$g =
0.67^{+0.18}_{-0.11}$)} and a significantly tapered gap edge (`round
wall'), but see no evidence that the inner disk, whose existence is
predicted by the
spectral energy distribution, shadows the outer regions of the disk
visible in our images.  We marginally confirm the existence of an 
offset between the disk center and the star along the line of nodes; 
however, the magnitude of this offset (\Rev{$x=27_{-20}^{+19}$}\,mas) is 
notably lower than that found in our earlier $H$-band images
\citep{paper1}.  Intriguingly, we
also find, at high significance, an offset of \Rev{$y=69^{+49}_{-25}$}\,mas
perpendicular to the line of nodes.  If confirmed by future observations,
this would imply a highly elliptical\,---\,or otherwise asymmetric\,---\,disk gap
with an effective eccentricity of $e \approx 0.3$.  Such asymmetry would
most likely be the result of dynamical sculpting by one or more unseen 
planets in the system.
Finally, we find that the bright arc of
scattered light we see in direct imaging observations originates from
the near side of the disk, and appears brighter than the far side 
because of strong forward scattering.
    }

   \keywords{circumstellar matter -- protoplanetary disks -- stars: pre-main sequence --
   			stars: individual: LkCa 15
			-- planets and satellites: formation -- techniques: high angular resolution}

   \maketitle
%



\section{Introduction}

As potential indicators of planetary companions, transitional disks offer the tantalizing possibility 
of observing planet formation in action.
They were initially identified as protoplanetary disks with a reduced near-infrared excess, indicating depleted inner regions \citep{strom89,calvet05}. Millimeter-wave imaging has confirmed that these disks indeed contain large inner holes or annular gaps \citep[e.g.,][]{andrews11}. However, optical and near-infrared images do not always reveal these holes \citep{dong12}, indicating different spatial distributions of micrometer and millimeter-sized dust grains (see \citealt{djo13}). In addition, stars continue to accrete substantial amounts of gas despite being cut off from the main gas reservoir in the outer disk \citep{ingleby13,bergin04}, increasing the complexity of the puzzle.


To understand the complex geometry of these disks, 
multi-wavelength imaging is necessary. Only a handful of these gaps have been imaged in the optical/near-infrared \Rev{\citep[e.g.,][]{mayama12,hashimoto12,quanz13,avenhaus13,boccaletti13,garufi13}}, due to their high contrast ratio and proximity to the host star. A promising method to overcome these hurdles is angular differential imaging \citep[ADI;][]{marois06}, which utilizes the natural field rotation of altitude-azimuth ground based telescopes to improve high-contrast imaging sensitivity. ADI efficiently reduces the impact of the stellar PSF wings, though at the cost of also imposing a non-negligible amount of self-subtraction for off-axis sources such as planets and disks. For point sources, this effect is easy to determine \citep[e.g.,][]{lafreniere07,acorns}, but for extended sources such as disks, the subtraction effects are non-trivial and can affect the apparent morphology of the source. Forward modeling is a powerful tool for interpreting these images and extracting the disk geometry \citep[e.g.,][]{thalmann11,milli12,thalmann13}.

One transitional disk that has been the focus of significant attention during the past few years is \object{LkCa 15}, 
a Sun-like host to a transitional disk with an inner hole the size of our Solar System \citep[$\sim$50~AU, e.g.,][]{pietu07}. Despite the large gap, LkCa 15 displays a significant near-infrared excess \citep{espaillat07} and residual millimeter emission from small orbital radii \citep{andrews11}, implying that an inner AU-sized, optically thick disk exists within the gap in addition to the outer disk \citep{espaillat08,mulders10}. Sparse aperture masking observations have revealed an extended structure within the gap, which has been interpreted as a possible accreting protoplanet \citep{kraus12}.

Recent simulations explore the structures induced into protoplanetary disks by individual embedded planets, and predict their observable signatures at near-infrared (NIR), mid-infrared, and sub-millimeter wavelengths \Rev{\citep{zhu11, dodson11, jangcondell13, djo13}}. \Citet{kraus12} note that their planet candidate cannot be the body sculpting the wide gap, suggesting that additional
planets may reside in the gap.

In 2010, we reported the first spatially resolved imaging of the LkCa~15 gap in the near infrared
\citep[][hereafter ``Paper I'']{paper1}. The observations confirmed the presence and size of the gap as being broadly consistent with that inferred by the \Rev{spectral energy distribution (SED)} and seen at longer wavelengths, but also indicated a possible offset of the gap center from the location of the central star. This implied an eccentric gap edge, the likes of which have been suggested as a possible indication of planets within the gap, shepherding the disk material through their gravitational influence \citep[e.g.,][]{kuchner03,quillen06}. The observations also revealed a strong brightness asymmetry between the Northern and Southern parts of the disk, which by itself could either be interpreted as preferential forward scattering from optically thin material at the near side of the disk, or reflection from an optically thick surface at the far side of the disk. 
The disk orientation derived by \citet{pietu07} on the basis of asymmetries in their millimeter interferometry data favors the former scenario, whereas \citet{jangcondell13} assume the latter scenario.

Overall, the very small angular scale of the disk and the prohibitive brightness contrast limited the amount of quantitative results that could be gleaned from the ADI images in \citetalias{paper1} directly, necessitating the acquisition of deeper high-contrast imaging data as well as a comprehensive forward-modeling effort as a next step.

Here we present four new epochs of near-infrared high-contrast imaging of the \Rev{LkCa~15} disk in the $K\sub{s}$ band. Due to their superior Strehl ratio, the $K\sub{s}$-band data offer cleaner disk images than the previously used $H$-band data.  Furthermore, the availability of several epochs of observation provides a more robust foundation for interpreting the disk morphology, through comparison of consistencies and scatter between the epochs. In order to extract quantitative results on the disk geometry from the imaging data, we generated an extensive parametric grid of model disks as described in \citet{mulders10}, calculated their appearance in scattered light, forward-modeled them through the ADI process, and compare them to the data using a $\chi^2$ metric.

In the following, we first describe our observations in detail, followed by a description of the data reduction procedure. We then discuss our modeling efforts to derive the properties of the disk and its inner hole, and subsequently show the results of this modeling. Finally, we discuss the implications of our results and present our conclusions.


\section{Observations}

\begin{table*}[tbh]
\caption{Summary of high-contrast observations of LkCa 15. Each of the five epochs 
of observation is
assigned an identifier (ID) used throughout this work. The table lists the epoch of 
observation, filter band, detector integration time (DIT) for a single exposure, 
number of exposures co-added per stored frame, number of usable frames, total 
integration time, full width of the field of view (FoV), range of parallactic 
angles covered, whether or not the target
star was saturated in the science images (sat.), the number of PSF reference stars, 
and the publication describing the 
data in detail. 
}
\label{t:epochs}
\begin{tabular}{@{}lllrrrrrrlrl@{}}
ID &  epoch		& band & DIT (s)  & co-adds & \# frames & time (min) & FoV & parall.\ angles ($^\circ$) & sat. & PSF stars & reference\\[1mm]
\hline\hline
H1 &  2009-12-26 & $H$ & 4.2  &  3   & 162  & 33.8   & 19\farcs5 & 102.4--255.7  & no & 1 & \citetalias{paper1}\\
\hline
K1 &  2010-11-15 & $K\sub{s}$ & 10.0 &  1   & 384  & 64.0  & 20\farcs5 &  89.9--263.0 & \textbf{yes} & 0 & this work\\
K2 &  2012-02-04 & $K\sub{s}$ & 5.0  &  2   & 173  & 28.8  & 5\farcs1   & 105.5--250.4 & \textbf{yes} & 0 & this work\\
K3 &  2012-11-04 & $K\sub{s}$ & 0.2  &  20  & 491  & 32.9  & 10\farcs2   &  94.9--260.5 & no 
& 2 & this work\\
K4 &  2012-11-06 & $K\sub{s}$ & 0.2  &  20  & 231  & 15.5  & 10\farcs2   & 116.3--262.3 & no 
& 1 & this work
\end{tabular}
\end{table*}

This work is based on four epochs of $K\sub{s}$-band high-contrast 
observations taken
with Gemini NIRI \citep{niri} from 2010 to 2012 (Gemini Science Programs 
GN-2010B-Q-93, GN-2011B-Q-36, GN-2012B-Q-94).  In all cases, the adaptive optics 
was used to deliver diffraction-limited images, and the image rotator was 
operated in pupil-tracking mode to enable data reduction with angular differential
imaging \citep[ADI;][]{marois06}.  The plate scale was 0\farcs020 per pixel. We 
hereafter refer to these four epochs of 
$K\sub{s}$-band observation as K1--K4.  Likewise, we assign the label H1 to the 
earlier $H$-band data we published in \citetalias{paper1}, which were taken on Subaru 
HiCIAO \citep{hiciao} at a plate scale of 0\farcs0095.  
An overview of all observations and their numerical parameters 
is provided in Table~\ref{t:epochs}.

The $K\sub{s}$-band observations largely share the same observation strategy and instrumental
setup, which renders them well-suited to a systematic analysis.  The only change in 
strategy was the adoption of short detector integration times (DIT) in runs K3 and K4, 
which allowed for unsaturated imaging of the host star LkCa 15 in the science data. 
For this purpose, the NIRI detector was operated in co-add readout mode to avoid 
large readout overheads.  Furthermore, the detector area to be read out was windowed
from the full 1024$^2$ pixels down to 256$^2$ (K2) and 512$^2$ pixels (K3, K4), since
the astrophysical area of interest for this study lies within a radius of $\sim$$1\arcsec$
from LkCa~15.

\begin{figure}[tbh]
\includegraphics[width=\linewidth]{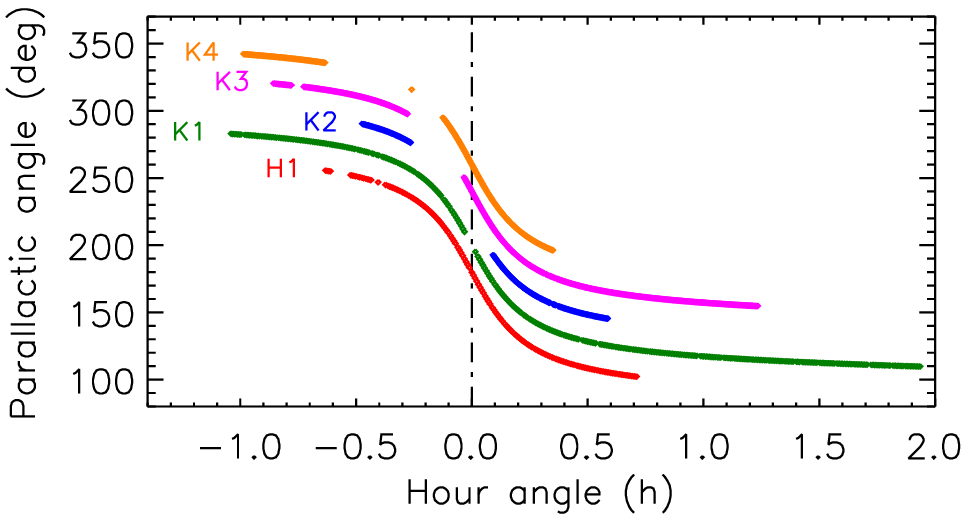}
\caption{Parallactic angle coverage of the five observing runs. The numbering of
    the vertical axis is accurate for epoch H1; all other epochs have been 
    translated upwards for visibility.}
\label{f:pa}
\end{figure}

Due to scheduling constraints and technical downtime, the observing runs 
vary in duration and continuity. The hour angle and parallactic
angle coverage is illustrated for each run in Figure~\ref{f:pa}.  While K1 comprises
the longest integration time and the fewest discontinuities among the $K\sub{s}$-band
runs, the target star is saturated in the science frames, which limits its usefulness
for accurate forward-modeling of the disk (cf.\ Section~\ref{s:modeling}). 
Furthermore, the field rotation stagnates
throughout the last hour of K1, thus its contribution to the ADI data reduction is
minimal.  Finally, the lack of unsaturated stellar PSFs in the science data reduces
the accuracy of forward-modeling for K1 and K2.  Overall, K3 provides the best 
combination of signal-to-noise ratio (S/N) and reliability amongst our observing 
runs.  We therefore adopt it as the benchmark dataset for our further analyses.

For the observing runs K3 and K4, shorter observations of two other target
stars were scheduled immediately before and after the science observations of LkCa~15,
to be used for point-spread function (PSF) reference.  These reference stars, 
\object{HD 283240} and \object{V1363 Tau}, were chosen so as to match LkCa~15 
closely in color, magnitude, and declinations in order to reproduce the PSF of the
science observations as well as possible.  However, V1363~Tau turned out to be a 
close binary, and thus it is unsuitable for a PSF reference.  Furthermore, vibrations in the
telescope (Andrew Stephens, p.c.)\ caused slight elongations of the PSF during those
epochs, which were less prominent for the reference stars than for LkCa 15.  As a 
result, we expect the reference PSFs to be of limited use for PSF subtraction of 
the science data.  These elongations should not have a significant impact on ADI-based
data reductions, though, since those techniques use the science observations  
themselves as PSF reference, and the position angle of the elongation is
stable with respect to the pupil.  Likewise, our forward-modeling analysis 
described in 
Section~\ref{s:modeling} takes this effect into account by convolving the model
disks with the actual PSFs of the science data.

\section{Data reduction}

\begin{figure*}[ptb]
\includegraphics[width=\linewidth]{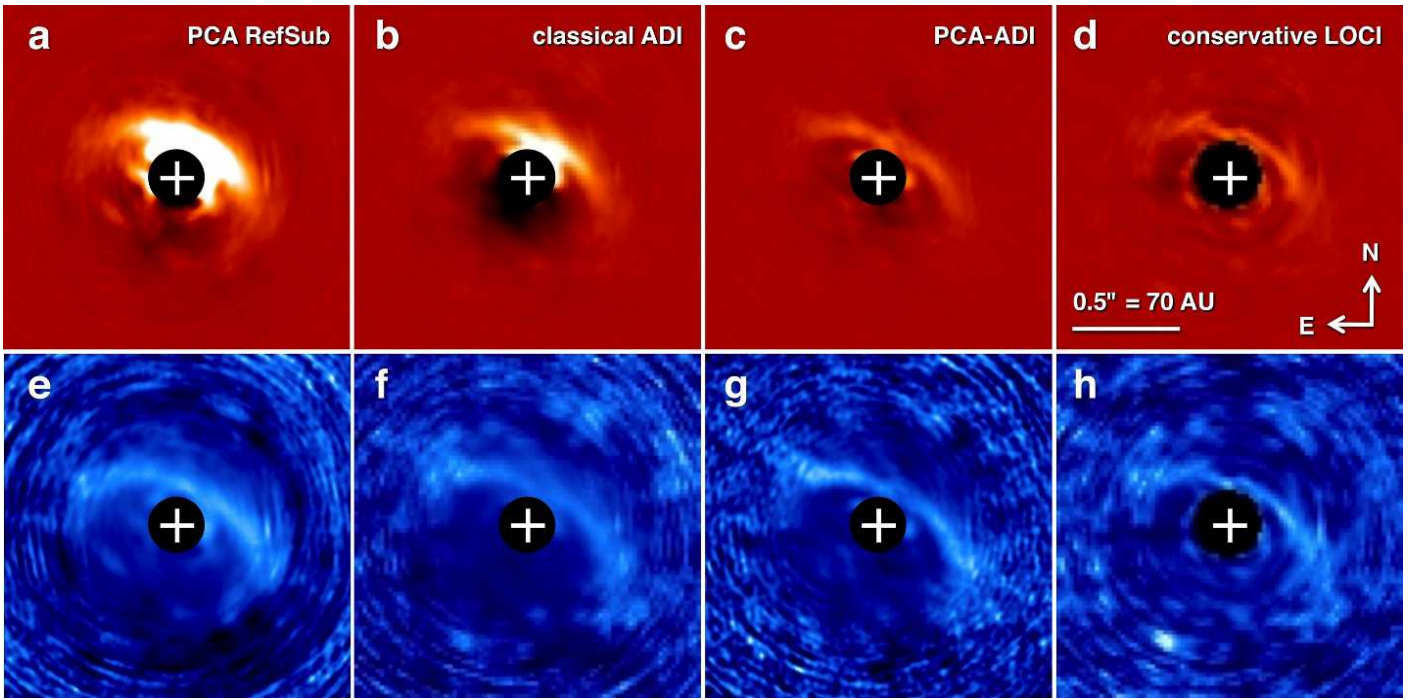}
\caption{Comparison of four high-contrast data reduction methods applied to the K3 dataset
	of LkCa~15. The top row \textbf{(a--d)} shows the resulting images at a fixed linear 
	stretch of
	$\pm1.6\times10^{-3}$ times the stellar peak flux. In order of decreasing conservation of 
	disk flux: \textbf{(a)} PCA-assisted reference PSF subtraction, \textbf{(b)}
	classical ADI, \textbf{(c)} PCA-ADI, \textbf{(d)} conservative LOCI.  
	The bottom row \textbf{(e--h)} shows the same images
	after renormalizing each concentric annulus around the star by the standard 
	deviation of the pixel values in the annulus, at a stretch of $\pm4\,\sigma$. The 
	resulting images resemble signal-to-noise (S/N) maps, though the 
	effective noise level is dominated by disk flux and thus overestimated in the inner
	$\sim$0\farcs5. Nevertheless, these images serve to reduce the dynamic range of the
	high-contrast images and visualize the characteristic crescent of positive disk flux 
	left behind by the differential imaging methods.
	}
\label{f:methods}
\end{figure*}

\begin{figure*}[ptb]
\includegraphics[width=\linewidth]{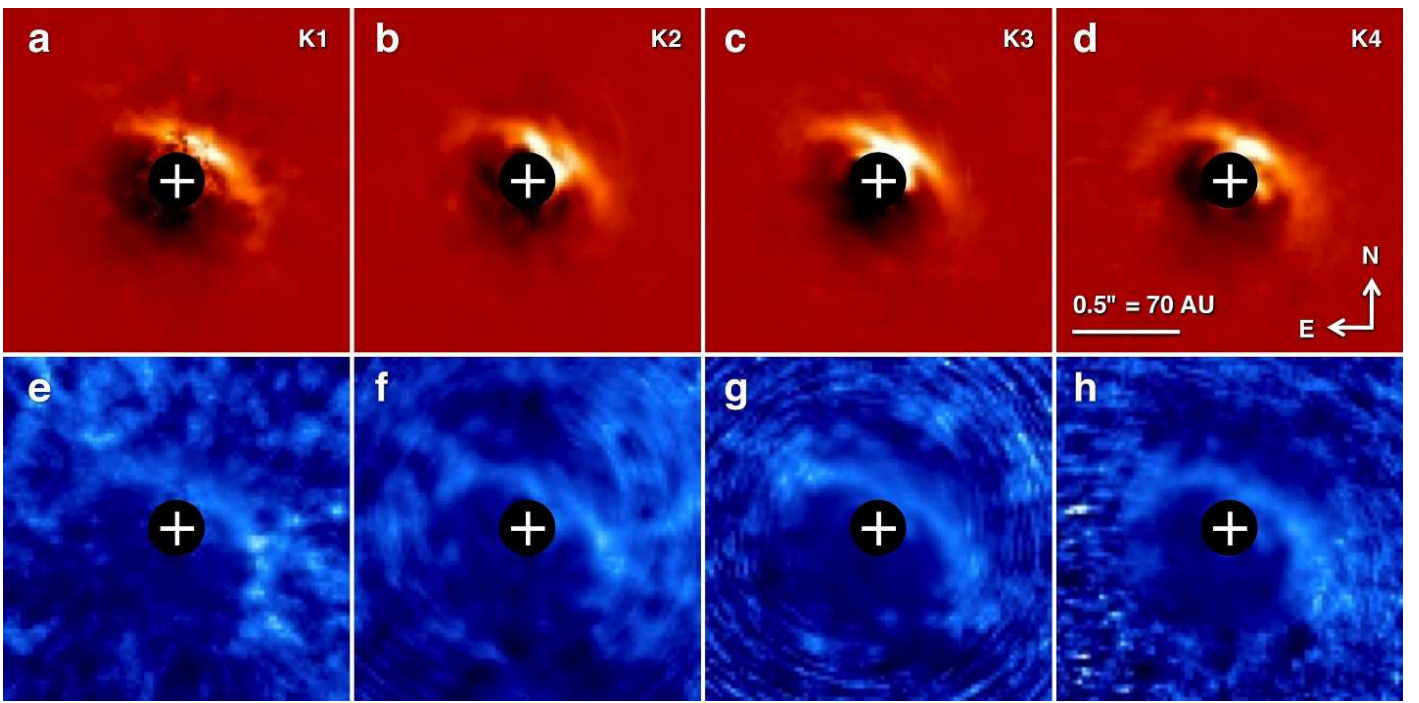}
\caption{Comparison of the four Gemini NIRI $K\sub{s}$-band observing runs of LkCa~15.
	All datasets were reduced with classical ADI. As in Figure~\ref{f:methods}, the 
	upper row shows the resulting images at a linear stretch of $\pm1.6\times10^{-3}$
	times the stellar peak flux, whereas the bottom row shows radially renormalized versions
	of the same images at a stretch of $\pm4\,\sigma$.
	\textbf{(a)} K1, \textbf{(b)} K2, \textbf{(c)} K3, \textbf{(d)} K4.  Since the 
	target star is saturated in runs K1 and K2, their flux normalization is 
	approximate.
	}
\label{f:epochs}
\end{figure*}

Despite the efforts of the adaptive optics system, the faint scattered light
from the LkCa~15 transitional disk is still 
overwhelmed by the diffraction halo
of its host stars in all datasets.  Differential imaging methods must be
employed to remove as much of the star's light as possible and recover the
disk flux.  Some of the disk flux is irretrievably lost in the process as
well, rendering the reconstruction of the disk's true appearance from the 
resulting images non-trivial \citep{thalmann11,milli12,thalmann13}. As
discussed in \citet{thalmann13}, the most rigorous way to infer physical
disk properties from such data is to generate a parametric grid of 
plausible numerical disk models, calculate scattered-light images from 
those models, and subject the theoretical disk models to the exact same data reduction process
as the science data.  The resulting differential images can then be
compared to those derived from the science data to constrain the range of
disk parameters consistent with observations.

A number of differential imaging techniques are available for use with ADI
data of circumstellar disks, offering a trade-off between effective 
suppression of the stellar halo (`aggressive' techniques) and conservation
of disk flux (`conservative' techniques).  The optimal choice of technique
depends on the disk geometry, the circumstances of the observation, and
the disk properties to be measured.

For the data at hand, we consider four differential imaging techniques, 
which we tested on the K3 dataset.  The resulting images are presented in
Figure~\ref{f:methods}. In order of increasing aggressiveness, the methods
are: 
\begin{description}
\item[\textbf{PSF reference subtraction}] based on
	principle component analysis \citep[PCA;][]{soummer12,pynpoint}. First,
	we used the Karhunen-Lo\`eve algorithm to decompose the dataset of the 
	PSF reference star into a mean PSF plus an orthonormal base of principal
	component images.  The image area within a radius of 6 pixels (0\farcs12) was 
	masked out to prevent the star's PSF core to dominate the definition of
	the principal components, as recommended in \citet{pynpoint}.  For each 
	frame in the science dataset, we then subtracted the mean reference PSF and
	matched the first
	$n$ principal component templates to the science frame by least absolute
	deviation fitting, and subtracted them out. The science frames were then
	derotated and co-added to produce the final image. We selected $n=2$ on the 
	basis of visual inspection of the results for $n=1\ldots9$.  
	
\item[\textbf{Classical ADI}] as presented in \citet{marois06}. An estimate of the
	stellar PSF is built by taking the mean of all science frames. The PSF is
	subtracted from all science frames, which are then derotated and co-added.
	We use \texttt{mean}-based frame collapse rather than \texttt{median} in 
	order to improve the residual noise characteristics, as recommended in
	\citet{acorns} 
	
\item[\textbf{PCA-ADI}] as employed by the \textsc{PynPoint} pipeline in 
	\citet{pynpoint}. This technique is essentially identical to the PCA-based
	PSF reference subtraction method described above, except that the science 
	dataset itself is used as the basis for the principal component templates
	rather than an external PSF reference star.  	Again, we selected $n=2$
	principal components to be subtracted.
	
\item[\textbf{Conservative LOCI}] as introduced in \citetalias{paper1} and
\citet{buenzli10}. The LOCI algorithm (Locally Optimized Combination of
Images; \citealt{lafreniere07}) is widely used in conjunction with ADI
for direct imaging of planets and brown-dwarf companions \citep[e.g.,][]
{marois08,thalmann09,lagrange10,carson13}. In its most common form,
it is too aggressive to image astrophysical sources beyond the size
of an isolated point source effectively; however, by choosing a stricter 
frame selection criterion \citepalias{paper1} or a larger optimization 
region area \citep{buenzli10}, the flux loss can be reduced far enough to
make the method viable for the imaging of circumstellar disks.  We refer to
this use of LOCI as ``conservative LOCI.''  For our $K\sub{s}$-band data, 
we find LOCI to be more aggressive than for the $H$-band data presented in 
\citetalias{paper1}---possibly due to the coarser plate scale---and 
therefore choose highly conservative LOCI parameters: a frame selection
criterion of $N_\delta=5$\,FWHM and an optimization area of 
$N_A = 10,000$ PSF footprints. See \citealt{lafreniere07} for detailed
description of these parameters.
\end{description}

\noindent
All data were flatfield-corrected and registered using Gaussian centroiding
of the target star prior to the application of differential imaging.  Since
all datasets are either unsaturated or mildly saturated, the expected 
registration accuracy is $\le$0.2\,pixel (4\,mas).

The dynamic range of the resulting images and the varying amounts of
flux loss among the four reduction methods render the images difficult to
compare directly.  For this purpose, we renormalize the images by dividing
the pixel values in each concentric annulus around the star by the 
standard deviation of those values.  The resulting images resemble S/N
maps, though the `noise' level is dominated by the disk signal, and thus
overestimated, out to $\sim$$0\farcs6$ (Figure~\ref{f:methods}e--h).
Nevertheless, they serve to illustrate the signal content of the 
differential images qualitatively.

The images from all four data reduction methods
paint a consistent picture of a crescent-shaped source of positive flux
visually consistent in shape, size, and orientation with the \Rev{$H$-band
image} published in \citetalias{paper1}, confirming
that we are indeed detecting scattered light from the surface of the
LkCa 15 pre-transitional disk.
Perhaps surprisingly, the signal-to-noise map derived from PSF reference 
subtraction (Figure~\ref{f:methods}e) does not differ fundamentally from
those made with ADI techniques (Figure~\ref{f:methods}f--h). Although
the reference subtraction method avoids self-subtraction of the disk, 
the net positive disk flux still causes the template-fitting routine to
overestimate and thus oversubtract the stellar halo in the science data.
Therefore, none of these images yield an unbiased view of the LkCa 15 
disk, making forward-modeling a necessity for quantitative analysis.

On the basis of Figure~\ref{f:methods}, \emph{we adopt classical ADI as our
differential imaging method of choice for the rest of this work}.  Due to
the saturation in epochs K1 and K2, PSF reference subtraction is ruled
out.  Of the remaining methods, classical ADI conserves the most disk flux,
is numerically transparent and linear, and requires the least 
computation time.  The latter point is relevant because the accuracy of a
forward-modeling analysis is limited by the sampling of the 
multi-dimensional model parameter space, and thus by the number of models
that can be evaluated in a reasonable timeframe.

Figure~\ref{f:epochs} shows the results of classical ADI applied to all four
four $K\sub{s}$-band datasets.  Overall, the crescent of scattered light
from the LkCa~15 transitional disk is consistently reproduced among the
epochs.  In particular, the position and orientation of the edge between
the disk gap and the bright side of the illuminated disk surface can be 
measured reliably.  The ansae of the disk gap are difficult to identify
due to oversubtraction and low S/N ratios, though, and the faint side of
the disk surface is not visible.


\section{Modeling}
\label{s:modeling}

\subsection{Experiment design}

Since the loss of disk flux in ADI image processing is irreversible,
the only robust method of extracting astrophysical information on the
LkCa 15 transitional disk from the data is forward modeling.  This means
generating a large number of plausible disk models by means of a 
radiative transfer code, simulating the observable appearance of those
disks in scattered-light imaging using a raytracing code, and finally
subjecting those images to the same ADI image processing that was 
applied to the science data.  The family of disk models that yield ADI
images consistent with those resulting from the science data can then be
used to derive the best-estimate values and confidence intervals for the
physical parameters of the disk.

Our forward-modeling setup comprises a total of nine independent free 
parameters:
\begin{itemize}
\item $r$, the radius of the disk's transitional gap outer edge, or `wall', represented by
	the scale factor $f := (56\,\mathrm{AU})/r$,
\item $i$, the inclination of its orbital plane, 
\item $g$, the Henyey-Greenstein forward-scattering 
efficiency of its dust grains, 
\item $w$, the `roundness' of the gap wall, 
\item $s$, the vertical \Rev{scale height} of the inner disk, \Rev{with $s=1$ corresponding
    to the canonical value derived from the SED (see Section~\ref{s:radtrans}),}
\item $o$, the orientation of the projected disk's rotation axis 
	(measured from North to East),
\item $c$, the flux contrast between the disk and the star, 
\item ($x, y$), the offsets of the disk's center from the
star parallel and perpendicular to the line of nodes, respectively. 
The line of nodes is the intersection of the inclined disk plane with the ``sky plane'' perpendicular to the line of sight; it defines the 
unforeshortened ``major axis'' of the projected disk.  The
parameter $y$ includes foreshortening, its observed value is 
$\cos(i)$ times the physical offset along the disk plane.
\end{itemize}

The radiative transfer code does not support eccentric disks.  However, 
since we discovered structure indicative of eccentricity in the LkCa~15
disk in \citetalias{paper1}, it is scientifically interesting to test
this hypothesis in the analysis at hand.  We therefore approximate
eccentric disks by calculating scattered-light images of azimuthally 
symmetric disks, translating them by a small offset $(x,y)$ with respect
to the star, and rescaling the brightness to take into account the new
center of illumination.  This is done in the third stage of our 
analysis (ADI processing).

\begin{figure*}[ptb]
\centerline{\includegraphics[width=0.85\textwidth]{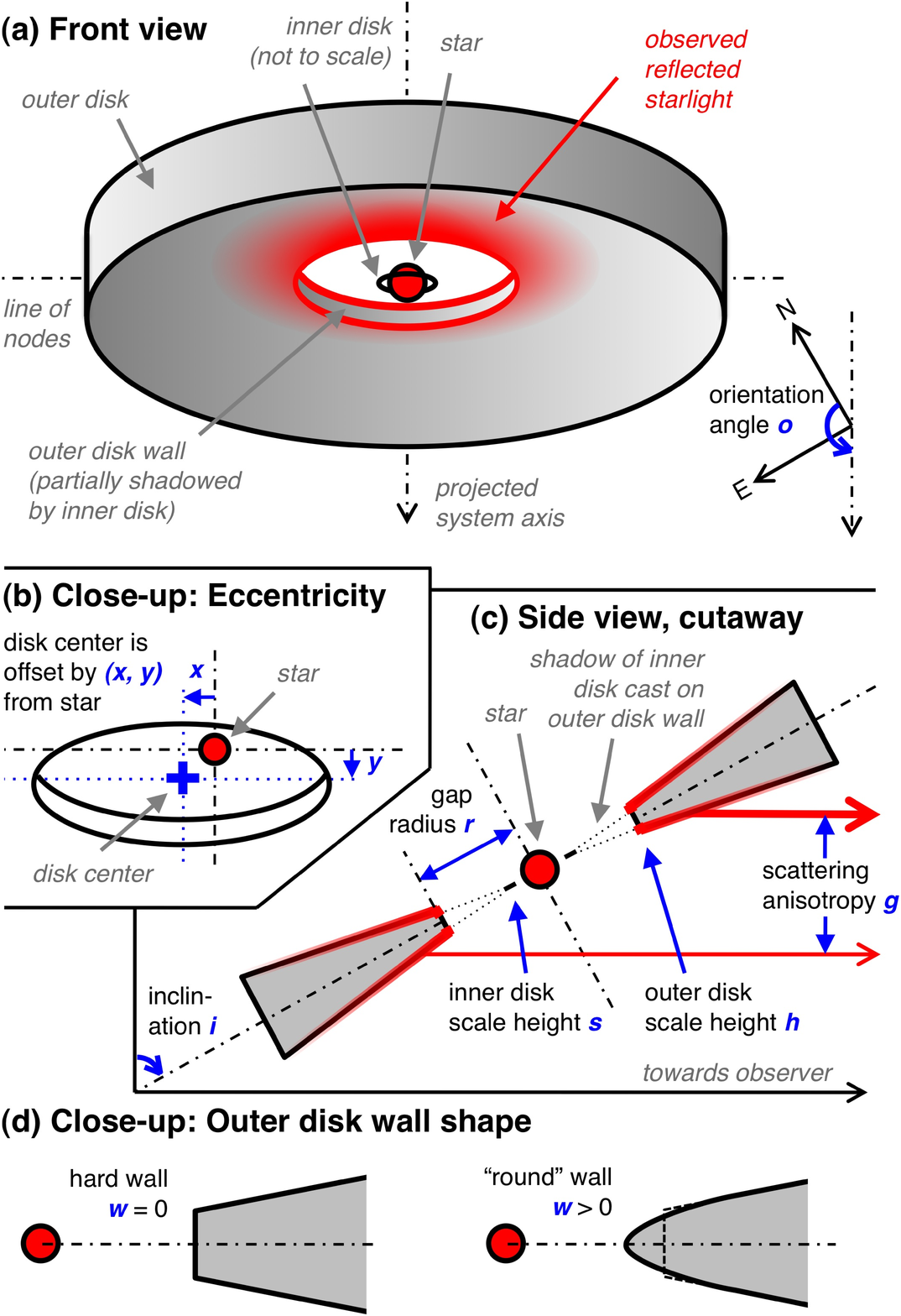}}
\caption{Sketch of the LkCa~15 system architecture as described by our model. Most of the bulk of the outer disk (shown in light grey) remains unseen; only the starlight reflected from its surface (shown in red) is observed with direct imaging.  Anisotropic scattering behavior greatly enhances the forward-scattered flux on the disk's near side, creating the bright crescent in the ADI images. The treatment of wall shape in our
model is visualized in more detail in Figure~\ref{f:sbp}. By default, the outer disk scale height $h$ is a fixed rather than a free parameter; see Section~\ref{s:y} and Appendix~\ref{a:h}. 
}
\label{f:infographic}
\end{figure*}

Figure~\ref{f:infographic} provides a graphical visualization of the
LkCa~15 system architecture as described by our model.  Note that the sketches are not to scale -- the SED predicts the inner truncation radius of the outer disk to be of order 50\,AU, whereas the inner disk is contained at sub-AU radii.  As a result, the inner disk is inaccessible to our direct imaging efforts, and can only impact our observations through the shadow it may cast on the wall of the outer disk.

The radiative transfer code used to generate self-consistent physical disk 
models is described in Section~\ref{s:radtrans}, the raytracing code
employed to generate scattered-light images from the disk models in
Section~\ref{s:scalite}, and the ADI processing and $\chi^2$ evaluation
in Section~\ref{s:formod}.

\subsection{Radiative transfer code}
\label{s:radtrans}
The radiative transfer code used in this paper is MCMax \citep{min09},
a disk modeling tool that performs 3D dust radiative transfer in a 2D
axisymmetric geometry. The code includes full anisotropic scattering
and polarization \citep{min12,mulders13a}, making it an ideal tool
for interpreting high-contrast images of protoplanetary disks. 

Besides radiative transfer, it solves for the vertical structure of
the disk using hydrostatic equilibrium and dust settling, yielding a
self-consistent density and temperature structure. The dust scale
height is in this case controlled by a scale factor $\Psi$, which
represents the reduction in scale height of the dust compared to that
of the gas. 

\begin{figure}[tbp]
\centerline{\includegraphics[width=\linewidth]{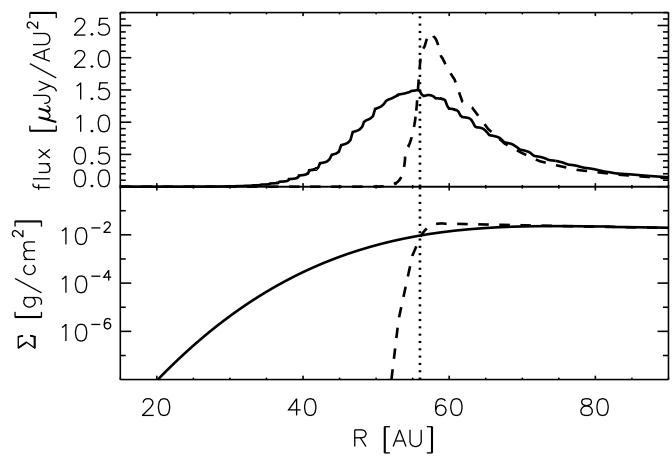}}
\caption{Azimuthally averaged surface brightness profile (top) and corresponding surface density profile (bottom), demonstrating the effect of rounding off the disk wall ($w>0$). The solid line shows the best-fit model with a wall shape of $w=0.30$, compared to a more vertical wall with $w=0.05$ (dashed line). The dotted line indicates the ``wall location'' $r$, corresponding to the radial peak in intensity.}	
\label{f:sbp}
\end{figure}

The model employed here is \Rev{based on} the optically \Rev{thick}
inner disk model described in 
\citet[][hereafter \citetalias{mulders10}]{mulders10}. 
The main modifications include:
\begin{description}
\item[\bf Anisotropic scattering.] The \citetalias{mulders10} model did not include
  scattering in the radiative transfer step of the simulation.
\item[\bf Dust properties.] We changed the dust size and composition to
  roughly reflect the observed brightness asymmetry, such that the
  imaging constraints do not significantly alter the SED fit. The dust
  properties are described in the next section and shown in Table \ref{tab:disk}.
\item[\bf Round wall.] Steep gradients in surface density are not a
  natural outcome of planet-disk interactions \citep{lubow06,crida07}. 
  Rather than a sudden increase at $r$, the surface density $\Sigma$ 
  gradually increases with radius up to a radius $R_{\rm exp}$, after which it follows 
  the surface density of the outer disk ($\Sigma_{\rm outer}~R^{-1}$).
  We use the exponential function described in \citet{mulders13b}, which is 
  characterized by a dimensionless width $w$ to described the spatial extent of this transition:
  \begin{equation}\label{eq:shape}
  \Sigma(R <   R_{\rm exp}) ~=~ \Sigma_{\rm outer}~R^{-1}~\exp\left( -\left(\frac{1-R/R_{\rm exp}}{w}\right)^3\right).
  \end{equation} 
  This smoothly increasing surface density gives rises to a rounded-off wall 
  in spatially resolved images, as shown in Figure \ref{f:sbp}.
  The wall extends further in than the anchor point for the exponential 
  turnover ($R_{\rm exp} \sim [1...3] r$), depending on wall roundness). 
  Instead, we use the radial peak in the intensity to trace the wall location $r$.
  Thus, $w$ represents the characteristic length scale of the exponential
  decay of the disk's surface brightness at the gap edge. It is a 
  dimensionless parameter normalized to the gap radius $r$. 
  Both the wall location $r$ and its roundness $w$ are free parameters.
\item[\bf Stellar properties.] We use the updated stellar parameters 
  from \citet{andrews11}, listed in \Rev{Table} \ref{tab:disk}.
\end{description}
 
The dust contains two components with different size distributions ($a_{\rm small}$ and $a_{\rm big}$) and different degrees of settling 
($s$ for the inner disk and $\Psi_{\rm small}$ and $\Psi_{\rm big}$ for the outer disk).
$\Psi_{\rm small}$ and $\Psi_{\rm big}$ are constrained by fitting the mid-infrared SED. $s$ determines the size of the shadow cast on the outer disk wall. Although fitting the near-infrared excess yields $s \simeq 1$ (\citetalias{mulders10}), we treat this parameter as free because the inner disk scale height is known to vary \Rev{\citep{espaillat11}}, and may be different for each observing epoch.

\begin{figure}[tbp]
\centerline{\includegraphics[width=\linewidth]{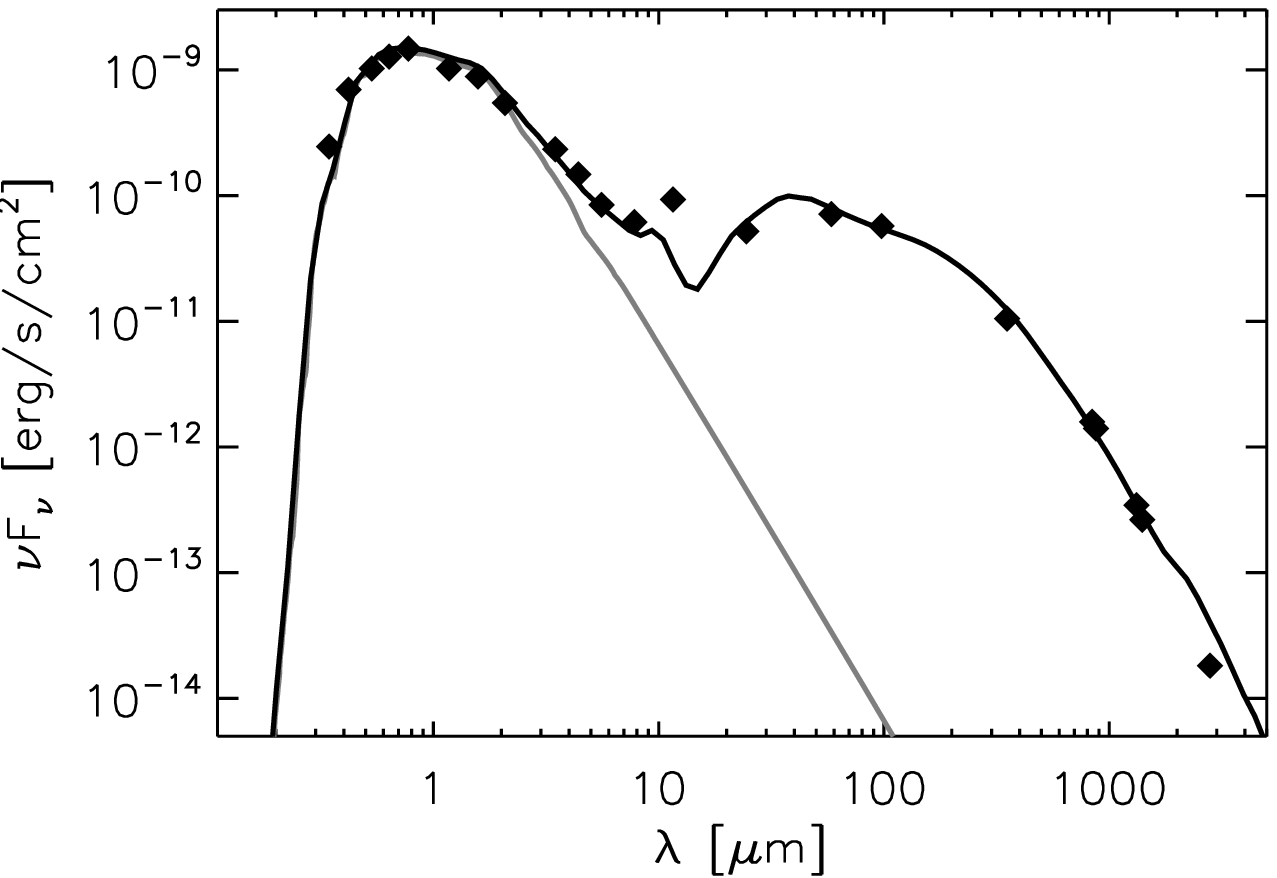}}
\caption{\Rev{Spectral energy distribution of LkCa15. Displayed with the solid line is the best fit disk model described in table 2. 
References – (UBVRI) \cite{kenyon95}; (JHK) \cite{skrutskie06};
Spitzer IRAC \citep{luhman10}; (IRAS) \cite{weaver92}; 
(submm) \cite{andrews05}; (mm) \cite{pietu06}.}}
\label{f:SED}
\end{figure}

\Rev{Since we updated both the stellar parameters and the dust composition since the \citetalias{mulders10}, it was needed to refit all parameters to the observed SED (figure~\ref{f:SED}.). All fit} parameters are shown in Table~\ref{tab:disk}. Note that parameters $s$, $w$ and $g$ are also used in the radiative transfer step, but are free parameters.

\begin{table}
  \title{Model parameters}
  \centering
  \begin{tabular}{ll}
    \hline\hline
    Parameter  & Value \\
    \hline
    $T_{\rm eff} ~[{\rm K}]$      &  4370  \\
    $L_* ~[L_\odot]$  &  1.22  \\
    $M_* ~[M_\odot]$  &  1.01  \\
    $d ~[{\rm pc}]$     & 140      \\ 
    \hline
    $R_{\rm in} ~[{\rm AU}]$  & 0.07 \\
    $R_{\rm out} ~[{\rm AU}]$  & 300\Rev{$^\dagger$} \\
    $R_{\rm gap} ~[{\rm AU}]$  & 4...56 \\
    \hline
    $M_{\rm dust} ~[M_\odot]$ & \Rev{0.01}  \\
    $M_{\rm dust,inner} ~[M_\odot]$ & $5 \cdot 10^{-11}$  \\
    \Rev{$p$} & \Rev{1$^\dagger$}\\
    \hline
    composition    & Solar$^\dagger$  \\
    $a_{\rm small} ~[\mu{\rm m}] $  & 0.1...1.5    \\ 
    $a_{\rm big} ~[\mu{\rm m}] $  & 100...500  \\ 
    $f_{\rm big,outer} ~[\%] $ & 99 \\  
    $ \Psi_{\rm small} $       & \Rev{0.65}   \\ 
    $ \Psi_{\rm big} $        & 0.2  \\ 
    \hline\hline
    \end{tabular}
    \caption{Disk parameters for the geometrical model. \Rev{Fixed parameters are indicated by a dagger ($\dagger$)}. The surface density profile is defined as $\Sigma(r)\propto r^{-p}$ and scaled to the total disk mass. The location of the gap edge is approximate due to its smooth nature, displayed here is the peak of the radial intensity (Fig \ref{f:sbp}). Opacities are calculated assuming a grain size distribution $f(a)\propto a^{-3.5}$.
      Solar composition:
      12\% MgFeSiO$_{4}$, 
      12\% MgFeSi$_{2}$O$_{6}$,
      12\% Mg$_{2}$SiO$_{4}$,
      12\% MgSi$_{2}$O$_{6}$, 
      15\% FeS, 
      40\% C. 
      Optical constants are from: Silicates \citep{1995A&A...300..503D,1996A&A...311..291H,1998A&A...333..188M}, Carbon \citep{1993A&A...279..577P}, Troilite \citep{1994ApJ...423L..71B}.
      }
\label{tab:disk}
\end{table}

\subsubsection{Grain size}

The absorption and scattering properties of the dust are set by the grain size,
structure and composition \citep[e.g.,][]{VanDerHulst}. The main
diagnostics of these properties in unpolarized images are the
wavelength-dependent albedo \citep{fukagawa10,mulders13a} and phase
function \citep{duchene04}. Because the latter is a highly non-linear
function of grain size, we use the Henyey-Greenstein phase function with assymetry parameter $g$
\citep{henyey41} to prescribe the phase function of our
particles, rather than varying the grain size. In doing so, we avoid
having to refit the SED for each different grain size, as the
grain properties can be kept fixed while varying $g$.

The grain composition is based on a condensation sequence for solar system
elemental abundances as described in \citet{min11}, see Table \ref{tab:disk}. We
assume the grains have an irregular shape, parametrized by a distribution of hollow spheres (DHS)
with a vacuum fraction ranging from 0 to 0.7 \citep{min08}. With
an MRN size distribution \citep{mathis77} from $0.1$ to $1.5$ micron, the albedo of these grains
is $\sim$0.5 in $K$-band with a phase function close to $g=0.5$.

\subsection{Scattered-light image simulation}
\label{s:scalite}

The scattered light images are calculated by integrating the formal
solution to the radiative transfer equations, using the density and temperature structure from the Monte Carlo simulation and the dust opacities as described above. The local scattered field is calculated in 3D using a Monte Carlo approach (see
\citealt{min12} for details). This scattered field includes a contribution
both from the star\footnote{The star is treated as a disk of uniform
 brightness with radius $R=1.65\,R_\odot$.} and from the disk, which
can be significant at near-infrared wavelengths. 

First, the images are calculated for a given inclination angle $i$ in spherical coordinates, preserving the radial grid refinement at the gap outer edge. Finally, this image is mapped onto a cartesian grid matching the pixel scale of the observations, yielding a scattered light image that can be further processed by ADI. 
By employing a scale factor $f$ for the field of view, we generate multiple cartesian images with the same resolution but different field of view from the same spherical coordinates image. Hence, a larger field of view (large $f$) corresponds to a smaller gap. This greatly decreasing runtime of the entire grid by skipping the radiative transfer and raytracing step that would otherwise be associated with changing the gap radius $R$. 

\subsection{Forward-modeling of ADI observations}
\label{s:formod}

\subsubsection{Theory}

The raw science data $\mathcal{I}$ used as input for the ADI process 
can be described as a sum of the stellar
PSFs $\mathcal{S}$ and the scattered-light images of the disk $\mathcal{D}$:
\begin{equation}
	\mathcal{I} = \mathcal{S} + \mathcal{D}.
\end{equation}
Since classical ADI using \texttt{mean}-based frame combination is 
linear \citep{marois06}, the resulting output image $\mathcal{O}$ produced by ADI 
processing of the imaging data $\mathcal{I}$ is equal to the sum of the 
ADI-processed stellar PSFs and the ADI-processed disk images:
\begin{equation}
	\mathcal{O} = \mathrm{ADI}(\mathcal{I}) = 
		\mathrm{ADI}(\mathcal{S}+\mathcal{D}) = \mathrm{ADI}(\mathcal{S}) + 
		\mathrm{ADI}(\mathcal{D}).  
\end{equation}
Since the stellar PSF remains largely static throughout the science data,
ADI effectively removes the bulk of the stellar flux, leaving behind a
halo of residual speckle noise.  Given good adaptive optics performance
(stable Strehl ratio) and enough field rotation (ideally across several
resolution elements at all considered radii), the noise is 
well-behaved and approximately Gaussian in concentric annuli around the star.
\begin{equation}
	\mathcal{O} = \mathrm{ADI}(\mathcal{D}) + \mathrm{noise}. 
\end{equation}
Thus, applying the ADI process to a noise-free model disk image
$\mathcal{D}_\mathrm{mod}$ yields an output image ADI($\mathcal{D}_\mathrm{mod}$)
that can be directly compared
with the resulting image from the science data. In the ideal case, 
subtracting the ADI-processed model image from the science data should
leave behind only pure noise:
\begin{eqnarray}
	\mathcal{O} - \mathrm{ADI}(\mathcal{D}_\mathrm{mod}) & =& 
		\mathrm{ADI}(\mathcal{D}) + \mathrm{noise}
	- \mathrm{ADI}(\mathcal{D}_\mathrm{mod}) \nonumber \\
	& = & [\mathrm{ADI}(\mathcal{D})-
	\mathrm{ADI}(\mathcal{D}_\mathrm{mod})] + \mathrm{noise} \nonumber \\
	& \mbeq & \mathrm{noise}.
\end{eqnarray}
Finally, assuming that the noise behaves in a Gaussian manner, one can
define a $\chi^2$ metric to measure the goodness of fit for a given 
disk model:
\begin{equation}
	\chi^2_\mathrm{mod} = \sum \frac{[\mathcal{O} - \mathrm{ADI}(\mathcal{D}_\mathrm{mod})]^2}{\sigma^2},
\label{e:chisq}
\end{equation}
where $\sigma$ is a map of the local standard deviation of the noise.

\subsubsection{Noise estimation}
\label{s:noiseestimation}

For highly inclined and well-resolved circumstellar disks, the radial noise
profile can be measured in sectors of the final science image unaffected by
the disk flux \citep[e.g.,][]{thalmann11, thalmann13}. However, the crescent
of scattered light from the LkCa~15 disk, as well as the surrounding 
oversubtraction regions, dominate our ADI images at all position angles, 
rendering it impossible to measure an unbiased noise profile.
During the 
exploratory phase of our experiment, we therefore started out with an 
estimated noise profile.  As soon as we had located well-fitting models
for each epoch, we iteratively redefined the noise profile for each dataset
as the 
standard deviation of the pixel values in concentric annuli measured in the
residual image $\mathcal{O}-\mathrm{ADI}(\mathcal{D}_\mathrm{best})$ for the best-fitting model
disk $\mathcal{D}_\mathrm{best}$. 
Such an \emph{a posteriori} definition of the
noise profile yields a reduced $\chi^2$ of $\sim$1 for the best-fit model
by design, and therefore cannot be used as an unbiased measure of 
absolute goodness of fit.  However, it provides a useful measure of
\emph{relative} goodness of fit, allowing us to converge on a best-fit set
of model parameters and to define confidence intervals around those 
parameters through a $\chi^2$ threshold.  

The noise profiles obtained with this method furthermore are used to 
generate more meaningful S/N maps than the ones shown in 
Figures~\ref{f:methods} and \ref{f:epochs}, since the disk flux no longer
dominates the definition of the noise profile.  The S/N map of the
residual image, $[\mathcal{O}-\mathrm{ADI}(\mathcal{D}_\mathrm{best})]/\sigma$, provides a
useful visual representation of the absolute goodness of fit, as well as a 
``sanity check'' on the assumed \emph{a posteriori} noise map.  Ideally,
it should look like random noise, with no coherent structure from the disk
left behind.  Similarly, the S/N map of the unsubtracted science ADI
image, $\mathcal{O}/\sigma$, provides a measure of the significance of the retrieved
disk signal, and thus of the scientific information content of the dataset.

\subsubsection{Implementation}
\label{s:implementation}

We perform the forward-modeling of ADI observations for our model disk images 
with a custom \texttt{IDL} code, which comprises the following steps:
\begin{itemize}
\item Load a simulated scattered-light disk model for a given set of 
	parameters $(r,i,g,w,s)$.
\item If $(x,y)$ are not zero, adjust the brightness distribution of the 
	image to represent illumination by the off-centered star.  For this 
	purpose, we multiply
	each pixel by the square of its distance from the image center, 
	then divide by the square of the distance from the new, offset position
	of the star.  When calculating distances, we assume that all disk flux 
	originates in the midplane; i.e., the projection effect from the plane's
	inclination is taken into account, but not the finite scale height of 
	the scattering disk surface.
\item For each exposure $n$ in the science data, generate a corresponding 
	`exposure' of the model disk image. The position angle of the model disk 
	in each frame is the input parameter $o$ plus the parallactic angle of
	the corresponding science exposure.  As part of the image rotation, the
	intended position of the star $(x,y)$ in the model image is mapped onto 
	the center of the model exposure.  This way, the model images need only be
	resampled once, minimizing aliasing effects.
\item Convolve each model exposure with a 15$\times$15 pixel image ($\approx$ 
    5 $\times$ 5 resolution elements) of the 
    unsaturated PSF core of the corresponding science exposure.  This 
    lends the simulated disk data the appropriate spatial resolution while 
    avoiding contamination of the model disk with the physical disk structure
    present in the science exposure at larger separations.   In the case
    of saturated science data (epochs K1, K2), we substitute a single
    external PSF for all exposures.
\item Perform classical ADI on the set of model exposures, yielding a
	noise-free ADI output image for the model disk.
\item Bin down both the model output image and the science output image 
	by a factor of 3 in both dimensions for the purpose of $\chi^2$
    evaluation.
	Since the diffraction-limited resolution $\lambda/D$ is 57\,mas = 2.8
	pixels, neighboring pixels are always strongly correlated. After the 
	binning, each pixel in the binned image can be treated as an independent
	data point. In practice, however, pixels may still be correlated to a
	certain degree due to large-scale structures remaining in the residual
	image.  We retain the unbinned images for visualization purposes.
\item Match the model output image to the science output image with a 
	least-squares optimization of the model output image's intensity scale
	factor.  We make use of the \texttt{linfit} function in 
    \texttt{IDL}, weighted with the \emph{a posteriori} noise map $\sigma$.
    This defines the final free parameter of the disk model, 
    the disk/star brightness contrast $c$.
\item We subtract the model output image from the science output image and
	calculate the $\chi^2$ value of the residual image as per 
	Equation~\ref{e:chisq}, using the \emph{a posteriori} noise map $\sigma$.
	We restrict the evaluation region to the annulus between the radii 
	0\farcs2 and 0\farcs8, thus excluding the center dominated by the stellar
	PSF core.  The evaluation region comprises $N=516$ binned pixels, each of 
	which roughly corresponds to a resolution element.
\end{itemize}

\subsubsection{Confidence intervals}
\label{s:confint}

The best-fit disk model is defined as the set of input parameters
yielding the minimum $\chi^2$ value, for each
epoch of observation.  We determine confidence intervals
around those parameters by characterizing the family of `well-fitting'
solutions bounded by a threshold of $\chi^2 \le \chi^2_\mathrm{min} + 
\Delta\chi^2$.  As in the case of our analysis of the HIP~79977 debris
disk \citep{thalmann13}, we find that the pixels in our residual images do
not have fully independent normal errors, as evidenced by traces of 
large-scale structure that remain visible in the S/N maps.  Applying the 
canonical $\chi^2$ thresholds would not take into account those remaining
systematic errors and thus overestimate the confidence in the best-fit
parameters.  We therefore choose a more conservative threshold of 
$\Delta\chi^2 = 32 \approx \!\sqrt{2N}$, which corresponds to a $1\,\sigma$
deviation of the $\chi^2$ distribution with $N=516$ degrees of freedom.
In other words, this threshold delimits the family of solutions whose residual
images are consistent with the \emph{a posteriori} noise map derived from
the best-fit residual at the $1\,\sigma$ level.

Under the assumption that the disk image does not change 
significantly with time, a global best-fit solution and well-fitting
solution family can be \Rev{obtained} by co-adding the four 
$\chi^2$ maps and adjusting the 
threshold $\Delta\chi^2$ by a factor of $\sqrt{4}=2$.  The assumption
is justified for most model parameters, since the disk gap is known to 
lie at a separation of $\sim$50\,AU \citep{espaillat08, andrews11}, 
which implies an orbital time scale of several hundred years.  \Rev{Thus,
we do not expect large-scale disk properties like the inclination, 
gap radius, or eccentricity of the outer disk to evolve measurably
between our observing epochs.}  On the other hand,
the inner component of the LkCa~15 pre-transitional disk orbits at
sub-AU separations and may therefore evolve at a time scale of months.
The model parameter $s$, which describes the vertical \Rev{scale height} of the
inner disk and thus the width of the shadow band on the outer disk 
wall, must therefore be considered variable between
K1, K2, and the pair of almost contemporary epochs (K3, K4).  
\Rev{In any
case, we present our results both for individual epochs and for the
combined analysis in order to allow direct comparison.  
As reported in Section~\ref{s:results}, no significant temporal 
variation is observed in any of the model parameters, including the
inner disk scale height $s$.  This validates the use of the global 
best-fit solution derived from all four epochs.}


\section{Results}
\label{s:results}

\subsection{Overview}
\label{s:resover}

In the early stages of this analysis, we explored parameter grids 
involving 2 or 3 model parameters at a time, keeping the other parameters
fixed, and iterated this process until a stable $\chi^2$ minimum was 
located for each observing epoch.  These \emph{preliminary best-fit} 
(PBF) solutions are listed in Appendix~\ref{a:pbf}.  While these models 
yielded visually convincing residuals and provided a necessary starting
point for further analyses, they lacked confidence intervals and were
not proven to be the global minima of the $\chi^2$ landscape.  We do not
document the specifics of these early searches in this work, since
they are rendered obsolete by the further analyses presented in the following
sections.  

The most robust way to establish confidence intervals 
is a comprehensive, finely-sampled exploration of
the 9-dimensional model parameter space.  
However, the radiative 
transfer simulation and the raytracing are computation-intensive procedures,
which renders such a brute-force approach impracticable.  
We therefore seek to reduce the dimensionality of the parameter space as 
much as possible beforehand.  

Simply exploring each parameter individually in one-dimensional analyses
around the best-fit solution is not a valid approach, since some parameters
are strongly covariant with each other.
For instance, the most striking feature of the ADI
images of LkCa 15 is the sharp transition from the dark disk gap to the
brightest edge of the reflected light on the near side.  For any given 
value of the gap radius $r$, there is only a narrow range of inclinations
$i$ that project the gap edge onto the correct apparent separation from 
the star to match the data.  However, a good match can be achieved over
a much wider range of inclinations if the radius is adjusted 
proportionally.  The well-fitting solution family therefore lies along
a diagonal in the $(r,i)$ plane, and is ill-described by its 
cross-sections with the $r$ and $i$ axes.

Nevertheless, we find that 3 parameters can be safely
decoupled from the full parameter space, and that one additional parameter
requires only minimal sampling.  This leaves only a manageable 5-dimensional 
parameter space to be explored thoroughly by brute-force calculations.

\begin{itemize}
\item $c$: The flux scaling of the model disk is set by least-squares 
	optimization using the \texttt{linfit} function of \texttt{IDL} as
	part of the evaluation code (cf.\ Section~\ref{s:implementation}).
	Thus, the disk/star contrast $c$ does not contribute a dimension to
	the parameter grid to be sampled.
\item $x, o$: The gap center offset $x$ along the line of nodes (the 
	``major'' axis of the projected gap ellipse) and the orientation 
	angle $o$ are the only two parameters whose effects on the disk image
	are not laterally symmetric.  Thus, we expect them to have well-defined
	optima largely independent of the other model parameters.  	
\item $s$: Our preliminary analyses surprisingly indicated that the 
	achievable $\chi^2$ monotonously decreases with decreasing inner disk 
	dust scale height $s$, reaching its minimum at $s=0$. $s=1$ corresponds
    to the hydrostatic equilibrium value of the inner disk scale height, 
    which is supported by observations \citep{espaillat08,espaillat10}.
	Fractional values of $s\ll 1$ correspond to a razor-thin disk, which 
    is not physically plausible.  The special case of $s=0$, on the other 
    hand, matches another physically plausible
	scenario: Either the inner disk is inclined with respect to the outer 
    disk, such that most of the inner disk's shadow bypasses the outer disk, 
    or the inner disk is radially optically thin as described in 
    \citetalias{mulders10}, in both cases leaving the outer disk fully illuminated. 
    We therefore limit the sampling in our analysis to the discrete values 
    $s=\{0, 0.5, 1\}$ to verify this result.
\end{itemize}


\subsection{Constraints on the orientation angle $o$ and on the center 
    offset along the line of nodes $x$}
\label{s:xo}

In order to determine the best-fit values and confidence intervals for
$x$ and $o$, we run 
three-dimensional parameter grids varying $(x,o,r)$ and keeping all
other parameters fixed at their preliminary best-fit values.  The
disk radius $r$ is included a grid parameter in order to allow the 
positions of the two ansae along the line of nodes ($\pm r + x$) to be
optimized independently of each other.	


\begin{figure}[tbp]
\centerline{\includegraphics[width=\linewidth]{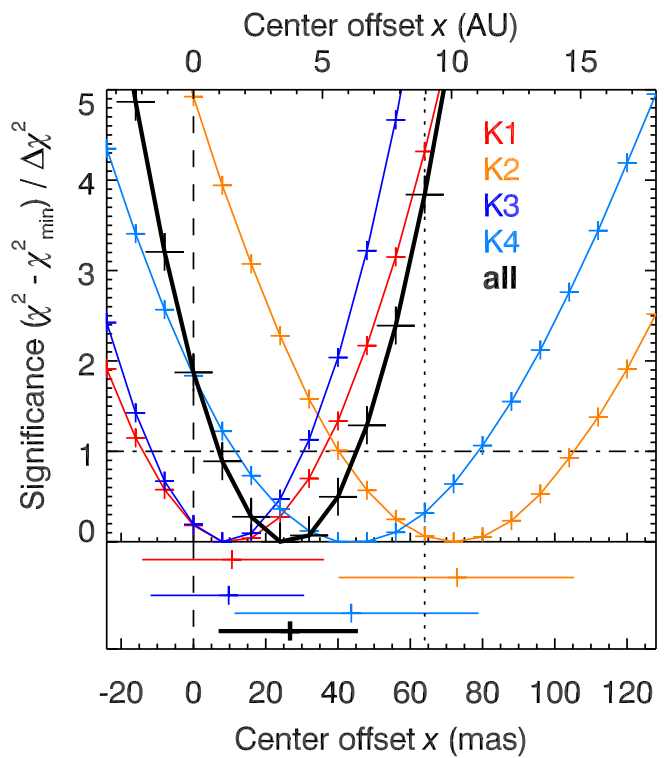}}
\caption{Constraints on disk center offset $x$ along the line of nodes.
	The plot shows the excess of the best-fit $\chi^2$ for a given $x$ 
	with respect to the minimal $\chi^2_\mathrm{min}$ achieved, normalized
	by the threshold value $\Delta\chi^2$.  The color-coded curves 
	represent each individual epoch, whereas the thick black curve is
	obtained by combining all four epochs.  
	The dashed line marks the case of a laterally symmetric disk ($x=0$), 
	whereas the dotted line marks the 
	tentative offset postulated in \citetalias{paper1}. Positive values
	of $x$ represent offsets in the direction of the western ansa.}	
	\Rev{The bottom panel shows the well-fitting range and the best-fit value
	for all five analyses.}
\label{f:x}
\end{figure}

\begin{figure}[tbp]
\centerline{\includegraphics[width=\linewidth]{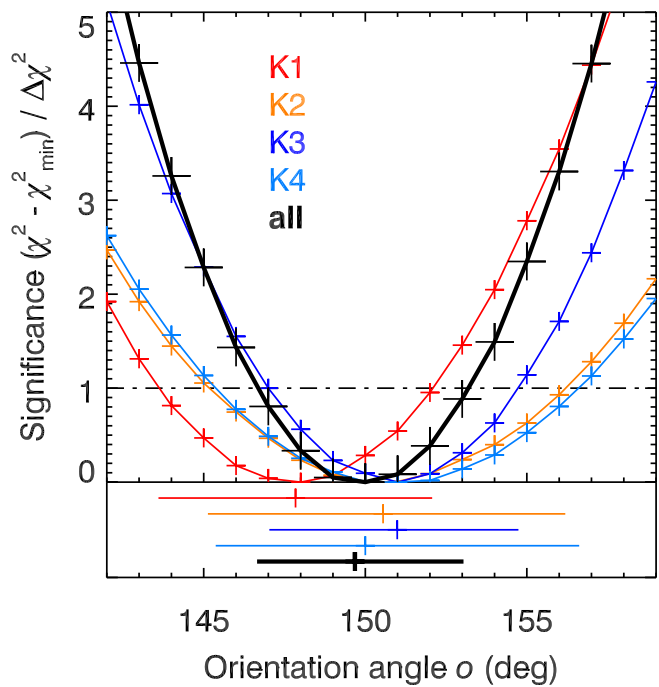}}
\caption{Constraints on the orientation angle $o$ of the disk's 
	rotation axis.
	The plot shows the excess of the best-fit $\chi^2$ for a given $o$ 
	with respect to the minimal $\chi^2_\mathrm{min}$ achieved, normalized
	by the threshold value $\Delta\chi^2$.  The color-coded curves 
	represent each individual epoch, whereas the thick black curve is
	obtained by combining all four epochs.  The angle $o$ is measured
	counter-clockwise from North to East.
	\Rev{The bottom panel shows the well-fitting range and the best-fit value
	for all five analyses.}}
\label{f:o}
\end{figure}

\begin{figure}[tbp]
\centerline{\includegraphics[width=0.8\linewidth]{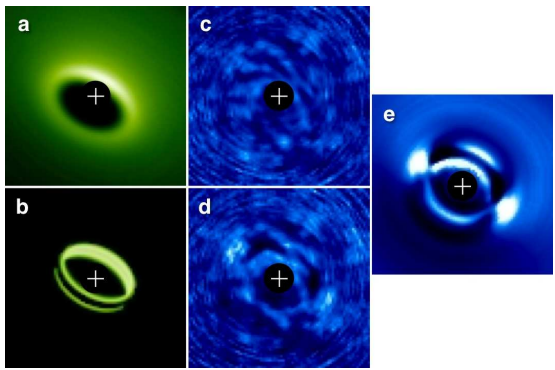}}
\caption{\Rev{Comparison of the best-fit solutions for the near-side
    scenario ($g=0.7$, $o=150^\circ$) and the far-side scenario
    ($g=0$, $o=230^\circ$) for the K3 dataset.
    \textbf{(a)} Image of the unconvolved best-fit model near-side
    disk model, at logarithmic stretch.  \textbf{(b)} The same for 
    the best-fit far-side model.  Note that in this configuration,
    the dark lane on the
    faint side is not generated by the shadow cast from the inner
    disk, but rather by the optically thick bulk of the outer disk
    blocking the view onto the illuminated face of the gap wall.
    The two parallel arcs of light are scattered light spilling
    around the edges of the near-side wall.
    \textbf{(c)} S/N map of the residual
    image after subtracting the ADI-processed near-side model image
    from the ADI-processed data, at a linear stretch of $\pm5\,\sigma$.
    \textbf{(d)}. The same for the far-side model. \textbf{(e)}. 
    Difference of the
    images (d) $-$ (c), at a linear stretch of $\pm2\,\sigma$. The
    $\chi^2$ of the far-side model exceeds that of the near-side model
    by 18\,$\Delta\chi^2$.  We therefore conclude that the bright 
    crescent in the data represents the strongly forward-scattering 
    near side of the disk.}}
\label{f:ocomp}
\end{figure}

Figures~\ref{f:x} and \ref{f:o} show the resulting best-fit values and
domains of the well-fitting solution families for $x$ and $o$, 
respectively.  The families from the four epochs appear as 
well-behaved $\chi^2$ curves with unique minima. While the $o$ values
from different epochs agree well, the $x$ values exhibit some scatter, in 
particular in the lower-quality epochs K2 and K4.

We find that the best-fit offsets $x$ consistently lie on the 
positive (western) side of the star in all four epoch, though the offsets
for the two high-quality datasets K1 and K3 are consistent with zero. 
Adding up the four $\chi^2$ maps yields global best-fit values and 
well-fitting intervals of $x = +24$ $[+7, +45]$\,mas 
(corresponding to $+3$ $[+1,+6]$\,AU in physical distance) and 
$o=150$ $[147, 153]$ degrees. A summary of all numerical results from
this analysis is included in Table~\ref{t:sanity}.

In \citetalias{paper1}, we postulated a tentative disk center offset 
of $x=+64$\,mas with a phenomenological error bar of $\pm6$\,mas on the 
basis of the H1 dataset.  This
error did not take into account the bias and systematic uncertainties 
from the ADI reduction, and therefore overestimated the accuracy of the
measurement.  However, the measured value is consistent with those of our
lower-quality datasets K2 and K4, whose S/N maps exhibit a similarly 
`patchy' structure
as that of H1.  The spread in $x$ values may therefore be due to a bias
caused by marginal lateral constraints on the gap; i.e., the optimization
of $x$ might be dominated by large-scale residual noise rather than
by the actual disk flux in the lower-quality datasets.  The fact that all
five measurements of $x$ have the same sign suggests an underlying
physical asymmetry that may be exaggerated by ADI under low S/N conditions.

Thus, overall, our \Rev{analysis} tentatively confirms the existence
of a positive disk center offset in $x$ direction, though at a 
significantly smaller magnitude than proposed in \citetalias{paper1}.

Note that for all our best-fit models, the bright crescent apparent
in the ADI images
is generated by the \emph{near} side of the disk, where the scattered 
light is greatly enhanced by anisotropic forward scattering 
\Rev{(`near-side scenario')}.  
This is in agreement with the predictions of \citet{pietu07} based on 
asymmetries in their millimeter interferometry images.  Furthermore, our
orientation angle agrees very well with their value of $o=150.7^\circ
\pm 0.4^\circ$.

\Rev{In \citepalias{paper1}, we demonstrated that} it is possible to 
make the far side of the disk rim brighter than the near
side within the parameter space of our model, \Rev{allowing for the 
`far-side scenario' as an alternative explanation for the bright
crescent in the ADI data.  In this scenario, the scattering anisotropy
is minimal ($g\approx0$), and the contrast is instead generated by
self-shadowing of the near-side gap wall, leaving only the illuminated
far-side wall exposed to imaging.
To explore this possibility, we have run a limited parameter grid
of models with $o=230^\circ$, $g=0$, coarse sampling of 
$s$ and $w$, and fine sampling of $i$, $f$, and $y$.  The latter three
parameters are most likely to influence the near-to-far side contrast.}

\Rev{The results are shown in Figure~\ref{f:ocomp}.  While the far-side 
scenario is capable of achieving a contrast comparable to the one in 
the near-side scenario, it requires the visible scattered light to be
tightly constrained to the gap wall.  The resulting image is a poor
match for the observed radially extended disk morphology.  The best
$\chi^2$ attained with the far-side model exceeds that of the near-side
model by 18\,$\Delta\chi^2$.  We therefore conclude that the bright 
crescent of scattered light in the LkCa~15 disk represents starlight
forward-scattered off the near-side surface of the outer disk.}


\subsection{Brute-force optimization and best-fit solutions}
\label{s:bestfit}

Having obtained best-fit solutions for $x$ and $o$, we perform a brute-force
exploration of the remaining parameter space, with fine sampling in 
$f, i, g, w, y$,  discrete sampling in $s$, and least-squares fitting in $c$. 
Table~\ref{t:sanity} defines the range and step size of the sampling in all 
parameter dimensions.

We find well-defined $\chi^2$ minima for all four epochs.  Table~\ref{t:sanity}
lists the model parameters corresponding to these best-fit solutions, as well
as the parameter ranges of the well-fitting solution family obtained with a 
threshold of $\chi^2 + \Delta\chi^2$ as defined in Section~\ref{s:confint}.  A
better characterization of the well-fitting solution family and the 
covariances between the model parameters is provided by the $\chi^2$ contour
plots in the following subsections.

\begin{table}[ptb]
\caption{Best-fit values, 1\,$\Delta\chi^2$ confidence ranges, and definition of the sampling grid
	for the LkCa 15 model parameters.}
\label{t:sanity}
\begin{center}
\begin{sideways}
\begin{tabular}{l||l||ll|ll|ll|ll||ll}
	& & \multicolumn{2}{c}{K1} & \multicolumn{2}{|c}{K2} &
	 \multicolumn{2}{|c}{K3} & \multicolumn{2}{|c||}{K4} & 
	 \multicolumn{2}{c}{\textbf{all epochs}}\\
Parameter & Search grid & best & range &  best & range & 
  best & range &  best & range &
  best & range\\[1mm]
\hline
Radius $r=56\,\mathrm{AU}/f$		& 43.1\ldots70.0	& 61.0 & [53.1, 67.7]
					& 54.8 & [44.7, 67.5]			& 54.8 & [49.5, 62.3]
					& 56.2 & [49.6, 63.0]			& 56.3 & [52.0, 62.6]\\
Inclination $i$ ($^\circ$)	& 28, 30, \ldots, 70		& 51	 & [46, 57]
					& 54 & [43, 62]					& 47 & [43, 53]
					& 44 & [36, 52]					& 50 & [44, 54]\\
Henyey-Greenstein $g$	& 0.45, 0.50, \ldots, 0.90	& 0.60 & [{\color{Gray} 0.45}, 0.76]
					& 0.86	& [0.67, {\color{Gray} 0.90}] 		& 0.72 & [0.55, {\color{Gray} 0.90}]
					& 0.60	& [{\color{Gray} 0.45}, 0.84]		& 0.67 & [0.56, 0.85]\\			
Wall shape $w$		& 0.05, 0.10, \ldots, 0.40		& 0.30 & [{\color{Gray} 0.05}, 0.38]
			& 0.31 & [{\color{Gray} 0.05}, {\color{Gray}0.40}]	& 0.27 & [{\color{Gray} 0.05}, 0.36]
					& 0.32 & [0.17, {\color{Gray}0.40}]			& 0.31 & [0.19, 0.36]\\
Inner disk settling $s$			& 0, 0.5, 1						& 0	& $\ll1$
					& 0	& $\ll1$						& 0	& $\ll1$
					& 0	& $\ll1$						& 0	& $\ll1$\\
Offset $x$ (mas)$^{(1)}$		& $-24$, $-16$, \ldots, $+128$		& $+11$ & [$-14$, $+36$]
                    & $+73$ & [$+40$, $+105$]		& $+10$ & [$-12$, $+31$]
					& $+44$ & [$+11$, $+79$]		& $+27$ & [$+7$, $+46$]\\	
---, (AU)			& $-3$, $-2$, \ldots, $+18$		& $+1$ & [$-2$, $+5$]
					& $+10$ & [$+6$, $+15$]			& $+1$ & [$-2$, $+4$]
					& $+6$ & [$+2$, $+11$]			& $+4$ & [$+1$, $+6$]\\
Offset $y$ (mas)		& $-140$, $-130$, \ldots, 0		& $-64$ & [$-111$, $-27$]
					& $-64$ & [{\color{Gray} $-140$}, $-6$]		& $-84$ & [$-131$, $-40$]
					& $-100$ & [{\color{Gray} $-140$},$-47$]		& $-69$ & [$-118$, $-44$]\\	
---, deprojected (AU)	& $-41$, $-39$, \ldots, 0	& $-14$ & [$-24$, $-6$]
					& $-14$ & [{\color{Gray} $-30$}, $-1$]			& $-18$ & [$-29$, $-9$]
					& $-22$ & [{\color{Gray} $-30$}, $-10$]			& $-15$ & [$-26$, $-10$]\\
Orientation $o$ ($^\circ$)$^{(1)}$		& 142, 143, \ldots, 159		&148 &[144, 152]
					& 151	& [145, 156]			& 151	& [147, 155]
					& 150  & [145, 157]			& 150 & [147, 153]\\ 
Flux ratio $c$	 (\%) 	& least-squares fitting		& 3.1$^{(2)}$ & [2.8, 3.6]
					& 2.9$^{(2)}$	& [2.5, 4.2]			& 2.6 & [2.2, 2.9]
					& 2.3 & [1.7, 2.5]			& ---\\[2mm]
\multicolumn{12}{p{22.7cm}}{\textsc{Notes.} --- (1) The parameters $x$ and $o$ were not
	sampled as part of the brute-force grid calculation, but in a separate analysis
	beforehand. (2) Disk/star contrast measurements from epochs K1 and K2 are unreliable, since
	the target star is saturated in the science data. \Rev{(3) Where the well-fitting ranges
	exceed the bounds of the sampled parameter grid, the grid bounds are listed 
	(in} {\color{Gray} gray color}\Rev{). Note that the range of the combination of all four
	epochs (last column) is fully enclosed in the grid.}
	}
\end{tabular}
\end{sideways}
\end{center}
\end{table}

\begin{figure*}[p]
\centerline{\includegraphics[width=0.85\linewidth]{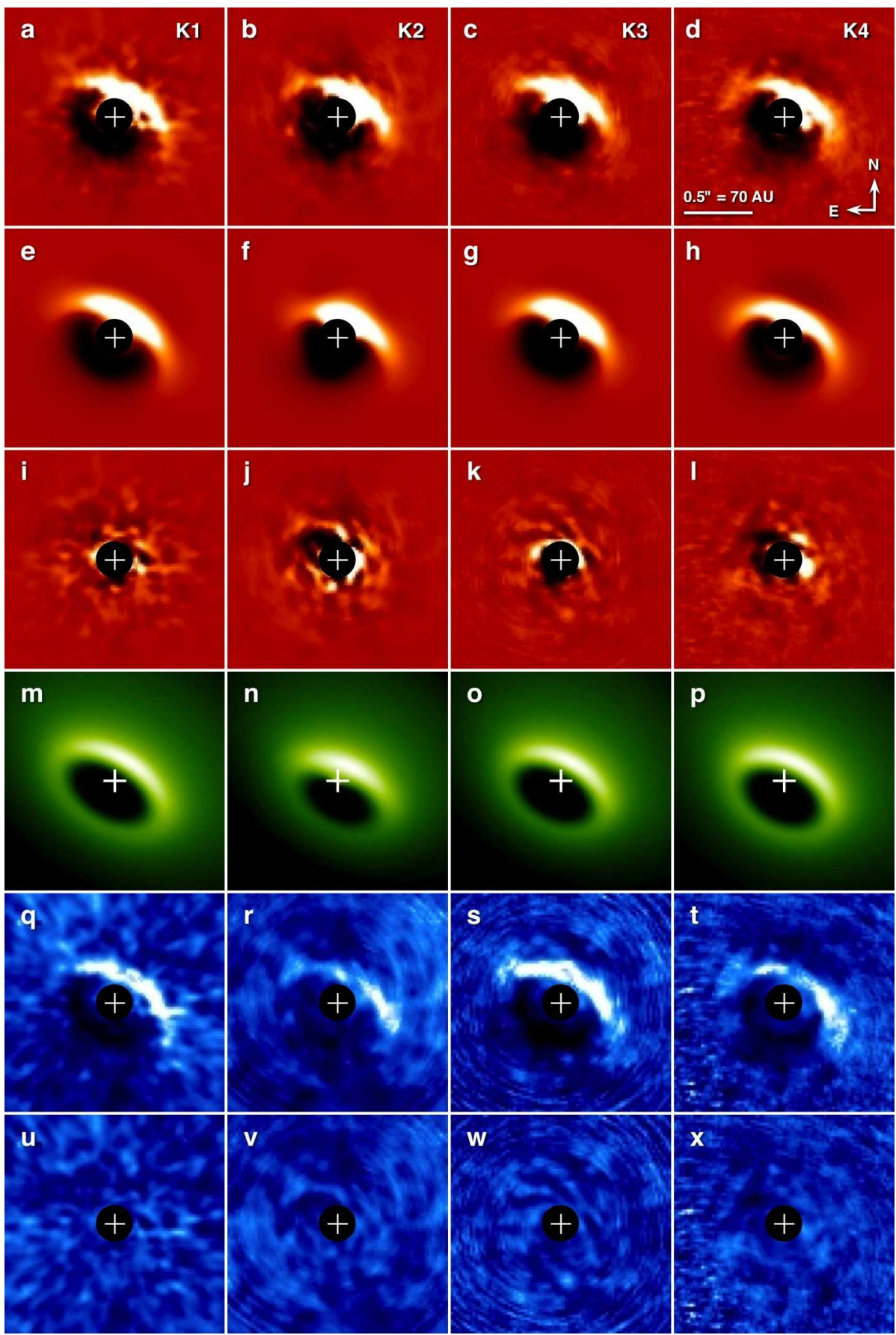}}
\caption{Best-fit models for the LkCa 15 disk. The columns represent
the four epochs of observation, K1--K4. \textbf{(a--d)} ADI output 
images from science data; \textbf{(e--h)} ADI output images from the
simulated model disks; \textbf{(i--l)} science--model subtraction 
residuals at the same linear stretch; \textbf{(m--p)} unprocessed
model disk 
images at logarithmic stretch; \textbf{(q--t)} S/N maps of the science
ADI output images at a linear stretch of $[-5\sigma,5\sigma]$; 
\textbf{(u--x)} S/N maps of the subtraction residuals at the same 
stretch.}
\label{f:bestfit}
\end{figure*}

Figure~\ref{f:bestfit} presents a visual overview of the best-fit model 
solutions and their agreement with the data for each epoch of observation.
The first row displays the classical ADI images derived from the data, 
while the second row shows the classical ADI images generated from the 
best-fit model disk images with the same set of parallactic angles.  The
third row represents the subtraction residuals of the first two rows, 
demonstrating that the overall morphology of the data is well explained
by the models.  The residuals are largely unstructured and well-behaved;
only in K2 a coherent arc reminiscent of the disk structure is retained.
There are no consistently recurrent features apparent in the four residual
images that would suggest observable disk structure beyond the scope of
the model parameter space, such as spirals, clumps, gaps, or planets.

The fourth row exhibits the unprocessed simulated scattered-light images
of the best-fit model disks in logarithmic stretch.  The disk 
architectures are roughly consistent between the epochs.  Note that the
position of the disk center is consistently offset from the star, bringing
the bright near side of the disk rim closer to the star.  In our model
parametrization, this corresponds to a negative value of $y$. This 
observation is further discussed below in Section~\ref{s:y}.

The fifth and sixth rows of Figure~\ref{f:bestfit} show the S/N maps
corresponding to the original ADI output images (first row) and 
subtraction residuals (third row), using the 
\emph{a posteriori} radial noise map derived from the residuals (third
row), as discussed in Section~\ref{s:noiseestimation}. 
  This reduces the dynamic range of the
images, allowing residual structures at various radii to be compared 
directly.  The images of the fifth row furthermore offer an estimate of
the overall information content of each observing run. Since the original
disk flux
can be assumed as approximately constant, the S/N at which the disk is 
detected at a given location is a measure of sensitivity and 
effectiveness of noise suppression.  

We can obtain a numerical measure
of the information content in each dataset by calculating the reduced
$\chi^2$ of the raw ADI images, which is defined as the RMS of the S/N
map divided by the number of pixels in the evaluation area.  The 
resulting numbers are given in Table~\ref{t:rchisq}.
\Rev{This confirms our choice of K3 as the most reliable dataset, and
explains the inferior performance of K2 in Figure~\ref{f:bestfit}.}

\begin{table}[htb]
\caption{Information content of the four datasets, as measured by the
reduced $\chi^2$ of the \emph{a posteriori} S/N map of the raw ADI
disk image. A higher reduced $\chi^2$ implies an ADI image that is
more inconsistent with being pure noise, i.e.,\ with more disk flux
measurable above the noise threshold.}
\label{t:rchisq}
\vspace*{-5mm}
\begin{center}
\begin{tabular}{l|rrrr}
	& K1 & K2 & K3 & K4\\
\hline
Reduced $\chi^2$	 & 3.9	& 2.1	& 5.4	& 2.5\\  
\end{tabular}
\end{center}
\vspace*{+5mm}
\end{table}


\subsection{Constraints on the wall shape parameter $w$}
\label{s:w}

\begin{figure}[tbp]
\centerline{\includegraphics[width=\linewidth]{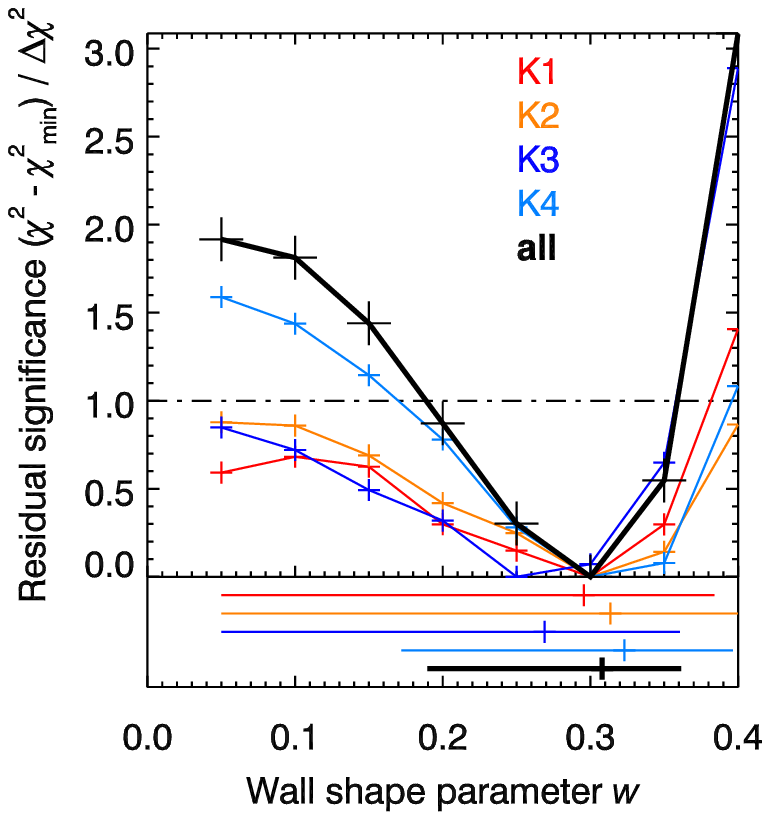}}
\caption{Constraints on the wall shape parameter $w$.
	The plot shows the excess of the best-fit $\chi^2$ for a given $w$ 
	with respect to the minimal $\chi^2_\mathrm{min}$ achieved, normalized
	by the threshold value $\Delta\chi^2$.  The color-coded curves 
	represent each individual epoch, whereas the thick black curve is
	obtained by combining all four epochs.  
	The bottom panel shows the well-fitting range and the best-fit value
	for all five analyses.
	The global best fit is 
	achieved for $w=0.31$, suggesting that the gap wall of the outer disk 
	is `fuzzy', extending over a significant range of radii rather 
	than being sharply constrained to a given radius ($w\approx0$). 
	\Rev{The bottom panel shows the well-fitting range and the best-fit value
	for all five analyses.}
	}
\label{f:w}
\end{figure}

Figure~\ref{f:w} presents the minimum $\chi^2$ achieved in our brute-force
grid search as a function of the gap wall shape parameter $w$.  
\Rev{The best-fit
solutions of all four epochs lie within the narrow interval of 
$w=[0.27, 0.31]$.}  While two individual epochs cannot exclude the 
possibility of an abrupt `vertical' wall ($w\approx0$), the combined
analysis yields a well-fitting
range of $w=[0.19, 0.36]$, clearly favoring an extended,
`fuzzy' radial transition from gap to outer disk.

\begin{figure}[tbp]
\centerline{\includegraphics[width=\linewidth]{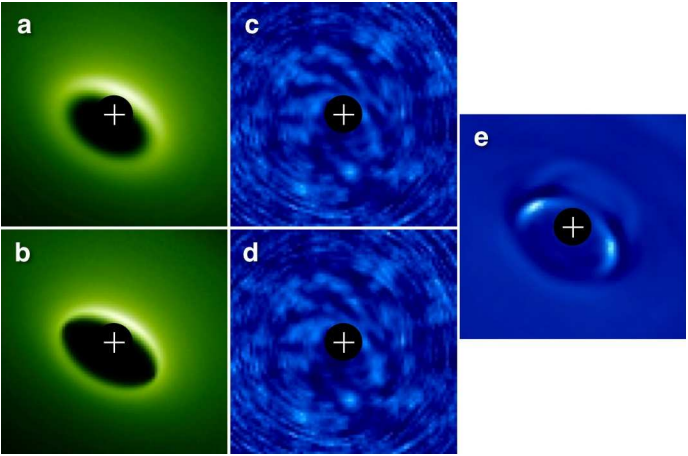}}
\caption{\Rev{Comparison of the best-fit solutions for $w=0.30$ and $w=0.05$
    for the K3 dataset.
    \textbf{(a)} Image of the unconvolved best-fit model disk with
    $w=0.30$, at logarithmic stretch.  \textbf{(b)} The same for 
    $w=0.05$.  \textbf{(c)} S/N map of the residual image after 
    subtracting the ADI-processed best-fit model image with $w=0.30$
    from the ADI-processed data, at a linear stretch of $\pm5\,\sigma$.
    \textbf{(d)}. The same for $w=0.05$. \textbf{(e)}. Difference of the
    images (d) $-$ (c), at a linear stretch of $\pm2\,\sigma$. The
    $\chi^2$ of the $w=0.05$ model exceeds that of the $w=0.30$ model by
    0.85\,$\Delta\chi^2$.}}
\label{f:wcomp}
\end{figure}

\Rev{Figure~\ref{f:wcomp} illustrates the difference between a round wall
$(s=0.30)$ and a vertical wall $(s=0.05)$ for the epoch K3.
We find no significant codependence of the best-fit value of $w$ with 
other model parameters.}

\subsection{Constraints on inner disk settling $s$}
\label{s:s}

\begin{figure}[tbp]
\centerline{\includegraphics[width=\linewidth]{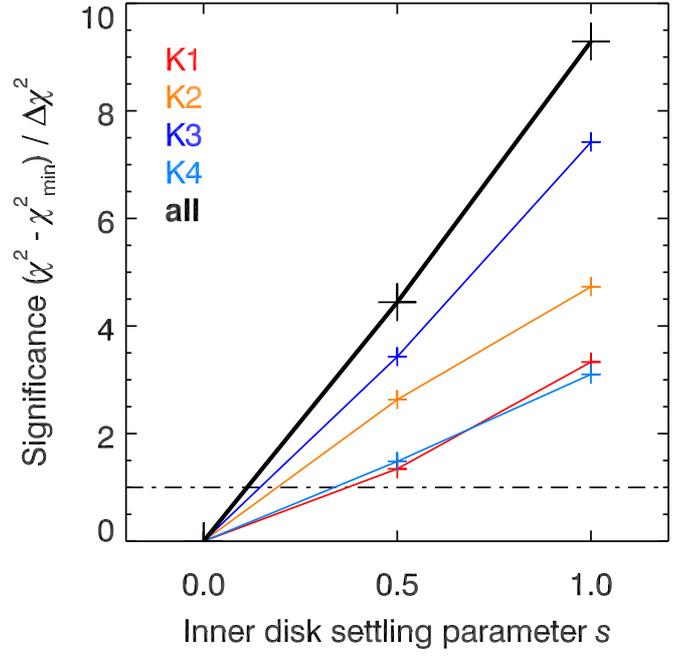}}
\caption{Constraints on the inner disk vertical settling parameter $s$.
	The plot shows the excess of the best-fit $\chi^2$ for a given $s$ 
	with respect to the minimal $\chi^2_\mathrm{min}$ achieved, normalized
	by the threshold value $\Delta\chi^2$.  The color-coded curves 
	represent each individual epoch, whereas the thick black curve is
	obtained by combining all four epochs.
	In all cases, the best fit is achieved with $s=0$, i.e., with a
	completely absent inner disk shadow.}
\label{f:s}
\end{figure}

\begin{figure}[tbp]
\centerline{\includegraphics[width=\linewidth]{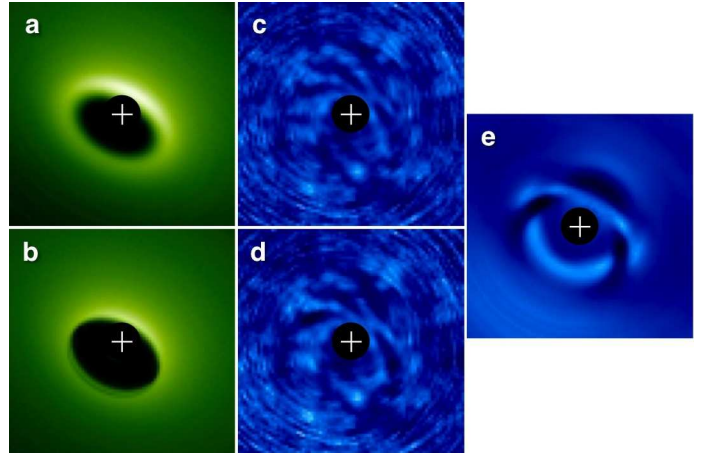}}
\caption{\Rev{Comparison of the best-fit solutions for $s=0$ and $s=1$
    for the K3 dataset.
    \textbf{(a)} Image of the unconvolved best-fit model disk with
    $s=0$, at logarithmic stretch.  \textbf{(b)} The same for 
    $s=1$.  Note the lane of shadowing from the inner disk visible
    on the far-side gap wall.
    \textbf{(c)} S/N map of the residual image after 
    subtracting the ADI-processed best-fit model image with $s=0$
    from the ADI-processed data, at a linear stretch of $\pm5\,\sigma$.
    \textbf{(d)}. The same for $s=1$. \textbf{(e)}. Difference of the
    images (d) $-$ (c), at a linear stretch of $\pm2\,\sigma$.  The
    $\chi^2$ of the $s=1$ model exceeds that of the $s=0$ model by
    7.4\,$\Delta\chi^2$.}}
\label{f:scomp}
\end{figure}

Figure~\ref{f:s} shows a plot of the minimal $\chi^2$ achieved in our
brute-force grid search for each of the three considered values of the
inner disk settling parameter $s=\{0, 0.5, 1\}$, in comparison to the
global $\chi^2$ minimum.  As expected on the basis of our preliminary
analyses (cf.\ Section~\ref{s:resover}), the global best fit is 
achieved with $s=0$ for all epochs,
whereas the best fits attainable with $s=1$ exceed the global $\chi^2$
minimum by \Rev{3--7}\,$\Delta\chi^2$ ($\Rev{10}\,\Delta\chi^2$ for all four epochs
combined).  

The figure suggests that the criterion for a well-fitting
solution ($\chi^2 \le \chi^2_\mathrm{min} + \Delta\chi^2$) is only 
achievable with $s \ll 1$. Taken at face value, this implies an 
unphysically small scale height for the inner disk, given that $s=1$ 
represents the expected hydrostatic equilibrium value that is 
compatible with the observed near-infared excess.  However, an
inner disk inclined with respect to the outer disk would leave
most of the outer disk wall unshadowed, and thus provide a physical
scenario for $s=0$.  We note that the alternative scenario of an 
optically thin dust halo in place of the inner disk, as 
proposed in \citet{espaillat07} and investigated in 
\citet{mulders10}, is also compatible with $s=0$, but the existence
of such a halo is difficult to justify from a physical point of view
(\citealt{espaillat10}, however, see \citealt{krijt11}).

\Rev{Figure~\ref{f:scomp} illustrates the differences between the 
best-fit solutions for $s=0$ and $s=1$.  The shadow of the inner
disk on the outer disk wall is visible as a dark ribbon in the
$s=1$ model image (Fig.~\ref{f:scomp}b).  The two illuminated 
edges of the wall are separated by slightly less than a resolution
element, and therefore cannot be distinguished in our data.
However, note that the effect of
shadowing from the inner disk in the $s=1$ case is not restricted
to the faint far-side gap wall, as one might expect.  While the
near-side wall would be self-shadowed in the case of a vertical
cutoff ($w=0$), its upper half remains exposed to direct view in
the case of a tapered disk edge ($w>0$), which is favored by our
analysis (cf.\ Section~\ref{s:w}).  Thus, the inner disk shadow
visibly truncates the inner edge of the tapered disk wall, 
hardening the gradient between the 
illuminated disk surface and the gap at all azimuths.  As a 
result, our analysis is more sensitive to the degree of shadowing
than the low S/N ratio of the far-side gap edge would suggest.}

Unlike all other model parameters, $s$ pertains to the inner rather
than the outer disk, which has an orbital time scale of months rather
than decades.  As a result, it is the only parameter for which one
may expect astrophysical variation from one observing epoch to the 
next.  However, we find consistent results for $s=0$ in all datasets.

We therefore conclude that, within the parameter space of our 
modeling effort, an inner disk inclined with respect to the outer
disk is the most likely scenario for LkCa 15.


\subsection{Constraints on the outer disk wall radius $r$}
\label{s:r}

\begin{figure}[tbp]
\centerline{\includegraphics[width=\linewidth]{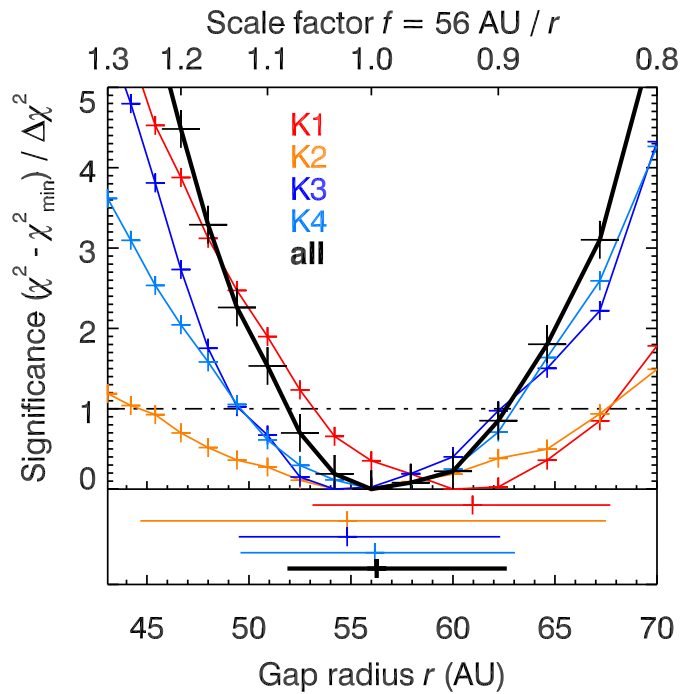}}
\caption{Constraints on the outer disk wall radius $r$, represented in
	our model by the scale factor $f=56\,\mathrm{AU}/r$.
	The plot shows the excess of the best-fit $\chi^2$ for a given $r$ 
	with respect to the global minimum $\chi^2_\mathrm{min}$, normalized
	by the threshold value $\Delta\chi^2$.  The color-coded curves 
	represent each individual epoch, whereas the thick black curve is
	obtained by combining all four epochs.
	\Rev{The bottom panel shows the well-fitting range and the best-fit value
	for all five analyses.}
	}
\label{f:r}
\end{figure}

Figure~\ref{f:r} shows the $\chi^2$ plot for the outer disk wall 
radius $r$.  The four epochs roughly agree with each other, with a
global best-fit value of $r=56$\,AU and well-fitting solutions for
the span of $r=[52,63]$\,AU.  This measurement is in excellent 
agreement with the value of $r=58$\,AU derived solely on the basis
of the SED by \citet{espaillat10}.

\subsection{Constraints on the inclination $i$}
\label{s:i}

\begin{figure}[tbp]
\centerline{\includegraphics[width=\linewidth]{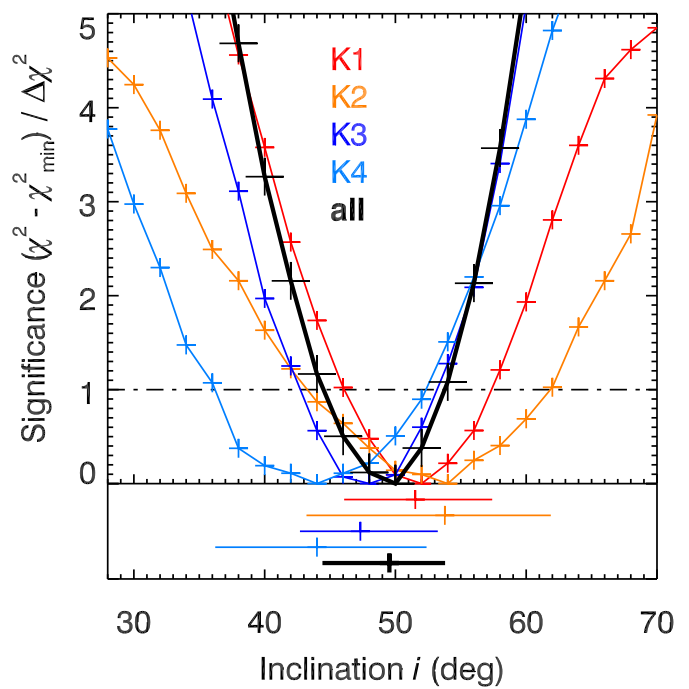}}
\caption{Constraints on the inclination $i$ of the disk plane.
	The plot shows the excess of the best-fit $\chi^2$ for a given $i$ 
	with respect to the global minimum $\chi^2_\mathrm{min}$, normalized
	by the threshold value $\Delta\chi^2$.  The color-coded curves 
	represent each individual epoch, whereas the thick black curve is
	obtained by combining all four epochs.
	\Rev{The bottom panel shows the well-fitting range and the best-fit value
	for all five analyses.}
	}
\label{f:i}
\end{figure}

Figure~\ref{f:i} presents the $\chi^2$ plot for the inclination $i$ of
the disk plane.  While there is some scatter among the lower-quality
epochs, the combination of all four epochs yields a very well-behaved
$\chi^2$ profile, with the best-fit solution at $i=50^\circ$ and 
well-fitting solutions spanning \Rev{$i=[44^\circ, 54^\circ]$}.  This is 
in excellent agreement with the $i=51.5^\circ\pm0.7^\circ$ inferred by \citet{pietu07}
from molecular line emission and confirmed by \citet{andrews11}, but 
notably higher than the $i=42^\circ$ assumed by \citet{espaillat10}.

\subsection{Constraints on the Henyey-Greenstein parameter $g$}
\label{s:g}

\begin{figure}[tbp]
\centerline{\includegraphics[width=\linewidth]{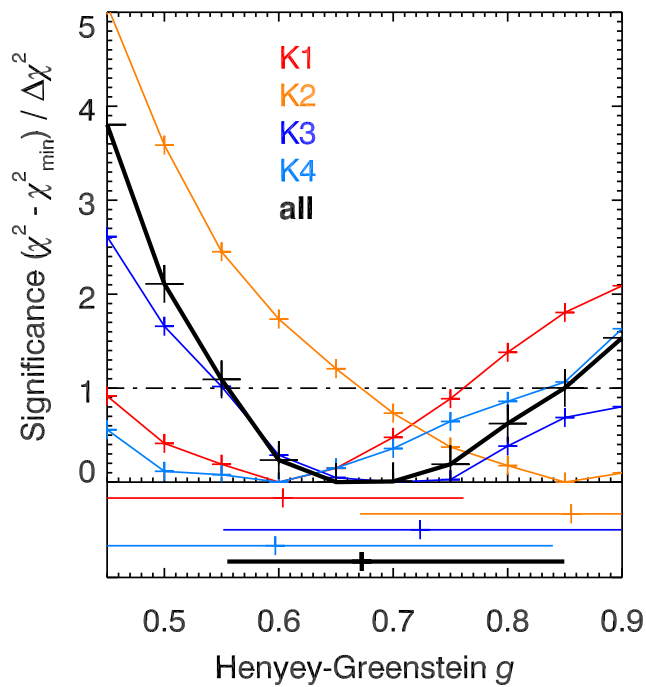}}
\caption{Constraints on the Henyey-Greenstein parameter $g$.
	The plot shows the excess of the best-fit $\chi^2$ for a given $g$ 
	with respect to the global minimum $\chi^2_\mathrm{min}$, normalized
	by the threshold value $\Delta\chi^2$.  The color-coded curves 
	represent each individual epoch, whereas the thick black curve is
	obtained by combining all four epochs. 
	\Rev{The bottom panel shows the well-fitting range and the best-fit value
	for all five analyses.}
	}
\label{f:g}
\end{figure}

Figure~\ref{f:g} displays the $\chi^2$ plot for the Henyey-Greenstein
parameter $g$, which describes the degree of anisotropic 
forward-scattering.  There is a considerable amount of diversity among
the epochs, with local best-fits ranging from \Rev{$g=0.60$} to 
\Rev{$g=0.86$}.
Since $g$ is a property of the size and shape distribution of the 
dust grains in the outer disk, it is not expected to exhibit any real 
astrophysical variation.  We therefore conclude that these variations
represent measuring uncertainy, and that the exact value of $g$ is 
difficult to pinpoint with our analysis approach.  Note that the 
Henyey-Greenstein formalism is a simplification of dust scattering
behavior, and may thus be an inaccurate description of the dust phase
function at hand.  \Rev{We do not consider values of $g>0.9$, since
the Henyey-Greenstein formalism becomes very inaccurate in that regime.}

Nevertheless, combining the $\chi^2$ curves of the four epochs yields
a clear best-fit solution at \Rev{$g=0.67$} and well-fitting solutions for 
the range of \Rev{$g=[0.56, 0.85]$}.  All of these values are notably on the
high end of the range expected for circumstellar disks, which explains
the extremely high contrast observed between the near and far sides
of the disk in scattered light (see also Figure \ref{f:hg}).

Because the system model is symmetric, one could argue that there is the possibility that the disk is oriented in the opposite way, i.e. the far side being the bright side, and the grains have a negative assymmetry parameter $g$. The phase function of aggregate particles can be backward scattering over a limited range of scattering angles \citep[see e.g.,][]{min10}. In our case, the scattering angles of the far side and the near side of the disk are approximately 134 and 34$^\circ$ respectively (given that the opening angle of the disk at the location of the wall is approximately 6$^\circ$ and the inclination is 50$^\circ$). To get the same contrast between these two scattering angles with a negative value of $g$, one would need \Rev{$g<-0.9$, which is highly unphysical}.

\subsection{Constraints on the center offset perpendicular to the line
	of nodes $y$}
\label{s:y}

\begin{figure}[tbp]
\centerline{\includegraphics[width=\linewidth]{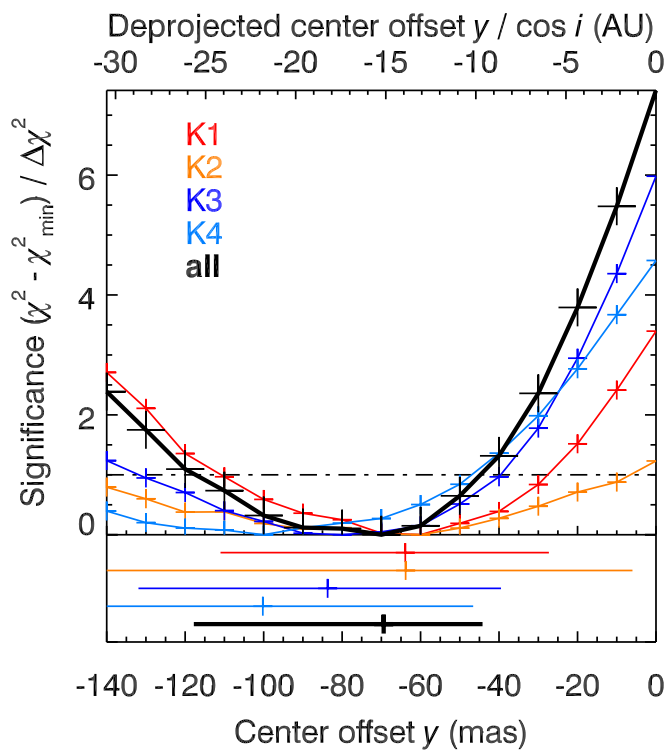}}
\caption{Constraints on the disk center offset $y$ perpendicular to
	the line of nodes.
	The plot shows the excess of the best-fit $\chi^2$ for a given $y$ 
	with respect to the global minimum $\chi^2_\mathrm{min}$, normalized
	by the threshold value $\Delta\chi^2$.  The color-coded curves 
	represent each individual epoch, whereas the thick black curve is
	obtained by combining all four epochs. 
	\Rev{The bottom panel shows the well-fitting range and the best-fit value
	for all five analyses.}
	}
\label{f:y}
\end{figure}

\begin{figure}[tbp]
\centerline{\includegraphics[width=\linewidth]{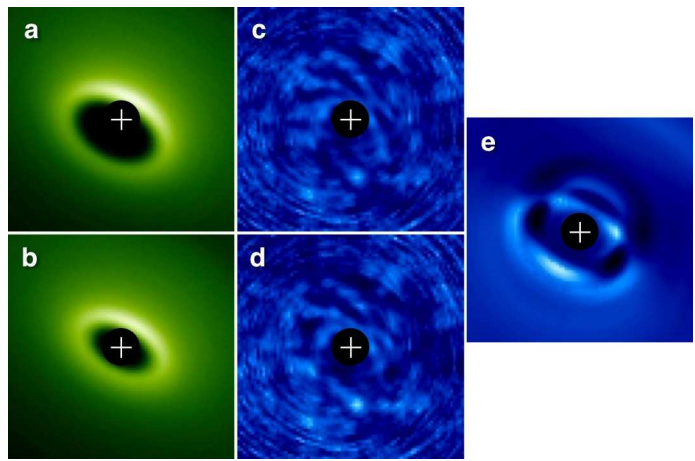}}
\caption{\Rev{Comparison of the best-fit solutions for $y=-80$\,mas and $y=0$\,mas
    for the K3 dataset.
    \textbf{(a)} Image of the unconvolved best-fit model disk with
    $y=-80$\,mas, at logarithmic stretch.  \textbf{(b)} The same for 
    $y=0$\,mas.  \textbf{(c)} S/N map of the residual image after 
    subtracting the ADI-processed best-fit model image with $y=-80$\,mas
    from the ADI-processed data, at a linear stretch of $\pm5\,\sigma$.
    \textbf{(d)}. The same for $y=0$\,mas. \textbf{(e)}. Difference of the
    images (d) $-$ (c), at a linear stretch of $\pm2\,\sigma$.  The
    $\chi^2$ of the $y=0$\,mas model exceeds that of the $y=-80$\,mas
    model by 6.0\,$\Delta\chi^2$.}}
\label{f:ycomp}
\end{figure}

Figure~\ref{f:y} shows the $\chi^2$ plot for the offset $y$ of the disk 
center from the star position perpendicular to the line of nodes.  The
four epochs agree well with each other.  Combining the $\chi^2$ curves
from all epochs yields a best-fit solution of \Rev{$y=-69$}\,mas, with 
well-fitting solutions for a range of \Rev{$y=[-118, -44]$}\,mas.  Accounting
for the foreshortening due to the inclined disk, this implies
a surprisingly large physical offset of \Rev{$y=15~[10, 26]$}\,AU along the 
disk plane.  Given the gap radius of $r\approx56$\,AU, this configuration
corresponds to an eccentricity of $e\approx0.3$.  \Rev{Figure~\ref{f:ycomp}
illustrates the difference between the eccentric best-fit solution for
the K3 data and the restricted best-fit solution with $y=0$\,mas.}

Since this study is the first to measure the disk center offset along
this dimension, its discovery is to be treated with caution until it can
be confirmed independently.  As Figure~\ref{f:y} demonstrates, our
combined $\chi^2$ analysis rejects the default
hypothesis of $y=0$ at the $7.5\Delta\chi^2$ level, thus there is little
doubt that a strong $y$ offset is necessary for a best-fit solution 
within our model grid.  However, we acknowledge the possibility that
this offset represents the model's effort to match a feature of disk 
architecture beyond the scope of its parameter space.

One degree of freedom that can conceivably be responsible for an 
apparent shift
of the disk image perpendicular to the line of nodes is the effective
scale height
of the \emph{outer} disk, represented in our model by the internal
parameter $\Psi_\mathrm{small}$.  As it increases, 
the upper and lower surfaces of
the disk move apart in the projected image.  Since only the upper surface
is seen in our reflected-light images, a change in $\Psi_\mathrm{small}$
may be perceived as a change in $y$ in our model.  

The outer disk scale height is not treated as a free parameter of
our model approach, since
its default value $\Psi_{\rm small}=0.5$ is constrained by 
the far-infrared SED \citep{mulders10}.  However, we ran a 
small-scale parameter analysis to determine whether its introduction
as a free model parameter would remove the need for a $y$ offset.
This analysis is described in Appendix~\ref{a:h}.  In order to keep 
the nomenclature consistent with the other free model parameters, in
particular with the inner disk scale height $s$, we named the new
free model parameter $h:=2\,\Psi_{\rm small}$. This defines the
default outer disk scale height as $h=1$, in analogy with the default
inner disk scale height $s=1$.

In the small-scale parameter analysis, we explored the extreme case 
of $h=2$ ($\Psi_{\rm small}=1$; i.e.,\ an inner disk scale height of
twice the expected value).
We found that the overall fit quality decreased significantly with 
respect to the default $h=1$ case, whereas the best-fit value
of $y$ did not change significantly from its $h=1$ best-fit value.  We
therefore rejected the use of $h$ as a free parameter and kept
$h=1$ fixed for our main analysis.  A physical offset of the disk center
from the star remains our best explanation for the observed disk images
at this point.

We note that our model implements an eccentric disk as a circular disk
shifted away from the star.  For low eccentricities $e\ll 1$, this method 
approximates an elliptical gap well.  At the observed eccentricity of
$e=0.3$, though, the approximation is quite inaccurate.  We therefore
acknowledge that our best-fit model likely cannot represent the disk's
true shape in all aspects, and that more elaborate disk models including
elliptical or even spiral-like disk architectures may yield improved 
fits in the future.


\subsection{Covariances between $r, i, g, w, y$}
\label{s:covar}

In Section~\ref{s:resover}, we opted for a brute-force parameter grid
search in the five model dimensions $r, i, g, w, y$, since we expected
that covariances between these parameters might render an independent
optimization of each individual parameter impractical.  Having 
calculated the $\chi^2$ values in this parameter grid for each epoch, 
we can now measure these covariances by mapping out the two-dimensional
contours of the well-fitting solution family for each pair of parameters.
For a pair of independent parameters, the contours ideally take the 
shape of a laterally symmetric ellipse, whereas a positive [negative]
covariance results in an ellipse visibly skewed towards the rising 
[falling] diagonal.

\begin{figure*}[tbp]
\centerline{\includegraphics[width=\linewidth]{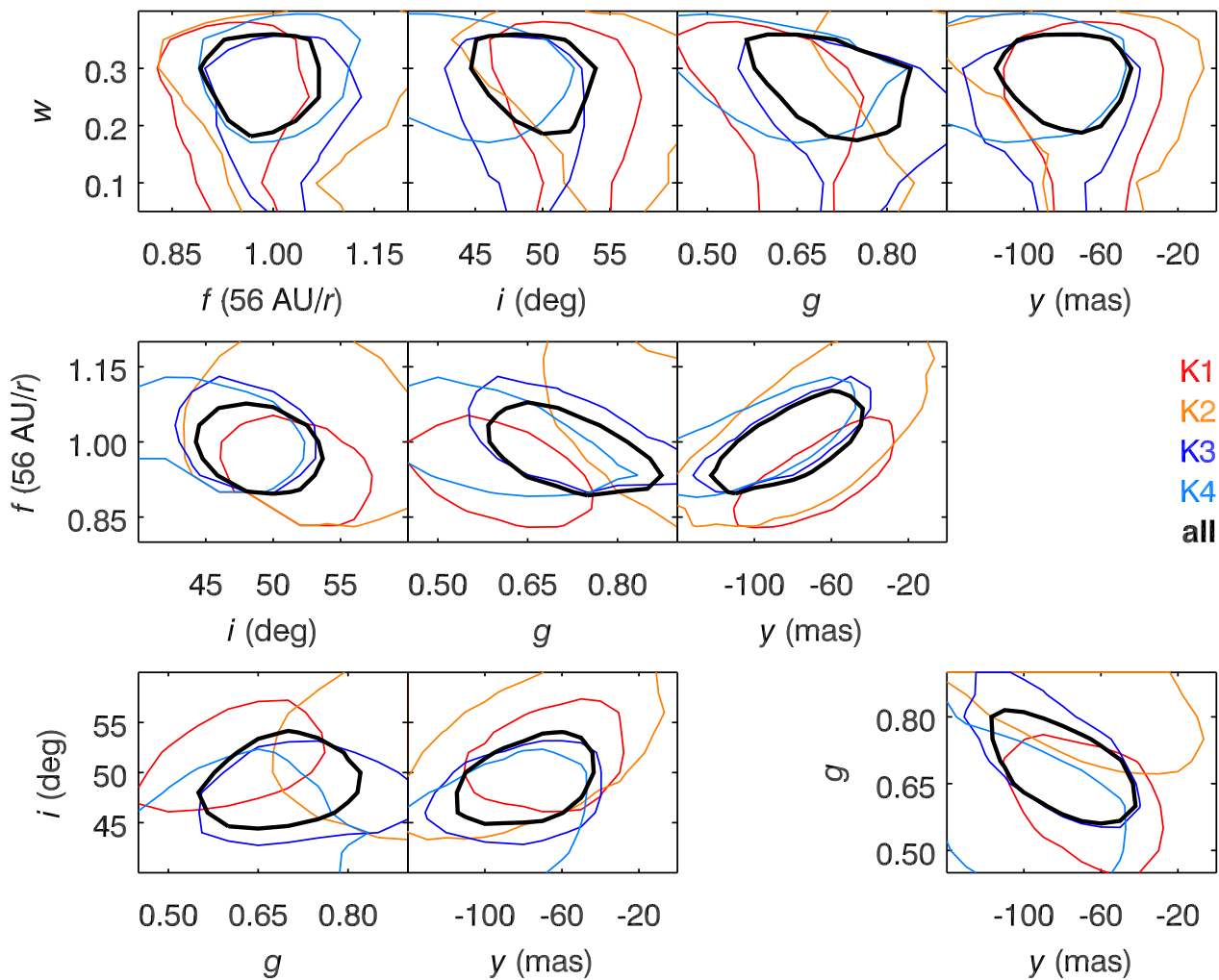}}
\caption{Visualization of the covariances between the model parameters
	$r, i, g, w, y$.  The contours delineate the well-fitting 
	solution family ($\chi^2 \le \chi^2_\mathrm{min} + \Delta\chi^2$) for each epoch,
	as well as for the combination of all four epochs.  The strongest
	covariances are seen between ($f,y$) and ($y,g$).
	}
\label{f:covar}
\end{figure*}

The results of this evaluation are presented in Figure~\ref{f:covar}.
Table~\ref{t:covar} lists the normalized correlation coefficients 
derived from the global well-fitting solution family contours for 
each pair of parameters.  Note that the contours only comprise a
limited number of grid points and are therefore only roughly 
accurate.

\begin{table}[h!]
\caption{Correlation coefficients between pairs of model parameters in
the brute-force grid-search study.}
\label{t:covar}
\centering
\begin{tabular}{l|rrrr}
	& $f$ & $i$ & $g$ & $y$\\
\hline
$w$	& \Rev{$-0.08$}	& \Rev{$-0.50$} 	& \Rev{$-0.50$}	& $-0.20$\\
$f$	& 			& \Rev{$-0.15$}	& \Rev{$-0.42$}	& \Rev{$+0.72$}\\
$i$	&			&			& $+0.12$	& \Rev{$+0.30$}\\
$g$	&			&			&			& $-0.55$\\
\end{tabular}
\end{table}

We find the strongest covariance between ($f,y$).  This positive
covariance implies a negative 
covariance between ($r,y$), given that $f:=56\,\mathrm{AU}/r$.
This behavior can be understood geometrically.  The most prominent
feature of the ADI images is the stark radial intensity gradient between
the
dark disk gap and the bright crescent of the forward-scattering near-side
rim of the
outer disk.  As $y$ is made more negative, the disk center is shifted 
further away from the star, bringing the disk's bright near side closer
to the star.  In order to keep the radial position of the gap edge
in its observed location, the disk radius $r$ must increase to 
compensate.  Decreasing the inclination $i$ provides another way
to widen the projected disk gap along the $y$ axis and thus restore the
position of the bright edge, but unlike a change in $r$, it 
changes the aspect ratio of the projected disk gap and thus degrades the
fit to the curvature of the imaged bright crescent.

The strong negative covariances between ($f,g$) and ($g,y$) are more
challenging to visualize. Decreasing $y$ and $f$ simultaneously, as 
discussed in the previous paragraph, should
result in an overall widening of the bright crescent in the ADI image,
even though its shape and location is roughly kept constant. This may then 
require an increase in forward-scattering efficiency $g$ to 
re-concentrate the flux in the central part of the crescent and thus 
restore the overall flux distribution.

In principle, observing the ansae and the far side of the gap edge would
break these degeneracies.  However, those features are not discernible in
our current datasets.  Next-generation high-contrast imaging facilities
like SPHERE ZIMPOL will likely be capable of such observations
\citep{thalmann08,djo13}.


\section{Discussion}

Overall, the disk architecture we derive from our near-infrared
imaging data agrees
well with the predictions from the SED \citep[e.g.,][]{espaillat08} as
well as millimeter and sub-millimeter interferometry \citep[e.g.,][]
{pietu07, andrews11}.  In the following subsections, we focus in 
particular on the
disk aspects to which these previous methods of observations were 
insensitive, but which have become accessible through high-contrast
imaging in reflected light.

\subsection{Disk orientation (near vs.\ far side)}
While the position angle of the projected system axis is easily 
measured in millimeter interferometry \citep{pietu07} as well as 
near-infrared imaging \citepalias{paper1}, it is more challenging to 
determine which of side of the disk is inclined towards the viewer.
\Citet{mulders10} proposed two possible architectures for the 
LkCa~15 system, both of which are compatible with the SED constraints,
but which predict opposite brightness asymmetries between the near
and far sides of the disk. In one scenario, the bright crescent 
observed in near-infrared imaging represents the directly illuminated
\emph{far side} of the gap wall, whereas the near-side wall is 
obscured by the bulk of the optically thick disk.  In the second 
scenario, the observed crescent instead represents light 
forward-scattered off the \emph{near side} of the outer disk rim, whose
observable flux is greatly enhanced over the far side by anisotropic
scattering.  The far-side scenario is qualitatively supported by a
disk modeling study by \citet{jangcondell13}, whereas \citet{pietu07}
and \citet{grady10} favor the near-side scenario based on asymmetries
in the millimeter interferometry data and in space-based \Rev{scattered-light}
images of the outer disk surface, respectively.  Our first 
near-infrared imaging analysis in \citetalias{paper1} did not allow us
to distinguish between these scenarios.

The quantitative approach presented in this work, however, clearly
favors the near-side scenario.  We achieve visually and numerically
convincing fits to the data using disk models with a high 
Henyey-Greenstein factor of \Rev{$g\approx0.67$}, which results in an
extreme enhancement of the near-side reflected light over the 
far-side flux.  Conversely, while we were also able to create an
opposite brightness asymmetry using the far-side scenario, \Rev{most of the 
scattered light is constrained to a narrow, self-shadowing gap wall, 
which fails to reproduce the radially extended morphology of the flux 
in our images (Figure~\ref{f:ocomp}).}

We therefore conclude that the Northwestern side of the LkCa~15 disk
is the near side, in agreement with \citet{pietu07} and
\citet{grady10}.

\subsection{Dust grain properties}

\begin{figure}[tbp]
\centerline{\includegraphics[width=\linewidth]{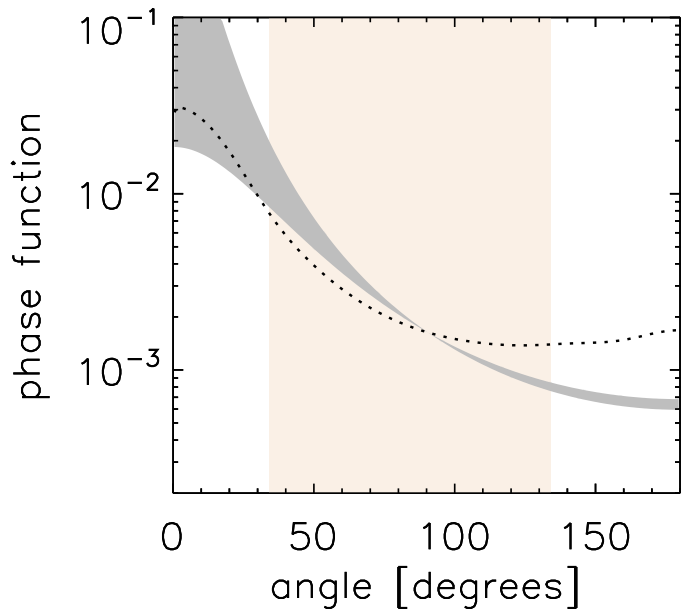}}
\caption{Phase functions of dust grains discussed in this work. Small angles are forward-scattering, large angles are backward-scattering. The gray region corresponds to the best fit range of the Henyey-Greenstein asymmetry parameter $g = [0.56, 0.81]$. The dotted line is the phase function of the dust grains used for the SED fit. The orange region indicates the observable range of angles, 34$^\circ$ to 134$^\circ$.}
\label{f:hg}
\end{figure}

The derived value of the Henyey-Greenstein anisotropic scattering parameter \Rev{$g\approx0.67~[0.56, 0.85]$} (cf.\ Fig.~\ref{f:g}) can be used to estimate 
the size of the particles in the upper atmosphere of the disk, which are responsible for the scattering. Note that the inclination and opening angle of the LkCa~15 disk constrain the observable phase function to the range of scattering angles roughly from 34$^\circ$ to 134$^\circ$ (Fig. \ref{f:hg}). For larger particles, the anisotropy typically increases, but most of this effect is concentrated at the smallest scattering angles (i.e.,\ the full phase function is no longer well represented with a Henyey-Greenstein function). At intermediate angles, the phase function usually flattens and, for very large aggregates, becomes locally backward scattering \citep[negative values of $g$; see e.g.,][and Min et al.\ in prep]{min10}. The rather large positive value of $g$ we find for LkCa~15 corresponds to particles that are roughly micron sized. This is in reasonable agreement with the grain sizes used in the upper atmosphere of our model disk, and it implies that these grains are not agglomerated in large aggregates.

\subsection{Spatial asymmetry}

In \citetalias{paper1}, we postulated a physical offset $x$ between the 
center of the LkCa~15 disk and the position of its host star along the line
of nodes, based on the lateral asymmetry of the $H$-band ADI image.  
In our model, we included two free
parameters $(x,y)$ to allow an arbitrary displacement of
the center of the circularly symmetric model disk from the star, where $x$
is measured along the line of nodes (the major axis of the projected 
disk gap) and $y$ perpendicular to it (along the minor axis of the
projected disk gap). 

Our analysis of the new $K_\mathrm{s}$-band observations yielded consistently
positive best-fit values of $x$ (cf.\ Fig.~\ref{f:x}), 
tentatively confirming the existence of a
lateral asymmetry in the appearance of the LkCa~15 disk.  However, the 
absolute values of $x$ show a considerable spread, with the lower-quality
epochs K2 and K4 supporting the large offset seen in \citetalias{paper1},
while the higher-quality epochs K1 and K3 advocate a much smaller offset
consistent with zero.  This may indicate that the real disk gap is roughly
centered along the $x$ direction, and that a feature further out on the
disk surface is responsible for the observed asymmetry.  For instance, a
spiral density wave extending beyond the western ansa could shift the 
perceived center of gravity of the disk flux towards that side, biasing
the analysis towards positive $x$ values in the lower-quality datasets.

In addition, our analysis suggests a previously unknown and suprisingly
large offset in the $y$ direction (cf.\ Fig.~\ref{f:y}).  
Both the sign and the absolute value
are consistent among all four $K_\mathrm{s}$-band epochs.  Correcting
for the foreshortening along the inclined disk plane, the best-fit offset
of \Rev{$-69$}\,mas translates to a physical displacement of $\sim$15\,AU, 
corresponding to $\sim$0.3 times the gap radius.  

This is a dramatic departure from the roughly symmetric disk architecture
that has been assumed in all previous studies of LkCa~15, and thus we
treat this result with caution until it can be independently confirmed.
However, we note that a sizeable disk asymmetry could well have gone 
undetected in previous observations.  Measurements of the system's SED,
as employed by \citet{espaillat08, espaillat10}, are inherently 
insensitive to purely spatial information.  While millimeter \citep{pietu07}
and submillimeter interferometry \citep{andrews11}
are capable of measuring disk information
on large spatial scales at high S/N ratios, the stellar photosphere is 
invisible at those wavelengths, rendering a direct comparison of the
disk and star positions impossible.  \Citet{andrews11} do not attempt to
fit any such offsets in their data, but note that their
disk center is $<70$\,mas from the expected stellar position based on 
absolute pointing accuracy. This is marginally consistent with our
best-fit offset.  We also note that the near side of the disk appears
brighter than the far side in \citet{andrews11}, which is qualitatively
expected if the near-side gap edge is closer to the star, and thus has
a higher equilibrium temperature.  Since millimeter interferometry traces
thermal emission rather than reflected light, anisotropic scattering does
account for this asymmetry.

Overall, our model treatment of the 
disk's eccentricity -- an azimuthally symmetric disk translated with
respect to the star along the disk plane -- is likely a simplification,
whereas the real disk may have an elliptical gap, spiral arms, or even
more irregular features (cf.\ HD~142527; \citealt{canovas13}).  
However, the fact that in our analysis all four epochs of observation
consistently suggests a significant offset is a strong indication of
an underlying physical asymmetry in the system architecture, regardless
of its detailed morphology.

\subsection{Wall shape}
To date -- and to our best knowledge -- the shape of a transitional
disk's gap wall has only been
detected in two other objects: HD 100546
\citep{panic13,mulders13b} and TW Hya (\citealt{ratzka07}, 
Menu et al. in prep.), both
with mid-infrared interferometry. In this work, we have inferred a
wall-shape parameter of $w\approx 0.30$ for LkCa~15, which describes
a gradual tapering of the outer disk surface density over a significant
range of radii (a `round' or `fuzzy' wall) rather than a hard 
vertical cut-off ($w=0$).  This value of $w$ is comparable to the wall
shapes described for the previously mentioned targets,
indicating that similar processes are likely at work shaping these walls. It should be noted, however, that steeper wall shapes in other objects
might have escaped detection as there may be less of a difference
observationally with the commonly assumed vertical wall. On the other 
hand, the missing cavities in the SEEDS survey \citep{dong12} might 
be interpreted as extremely extended round walls. 

A tapered surface density profile for LkCa~15 was also proposed by \cite{isella09}, as a way of reconciling the millimeter flux deficit with the existence of a near-infrared excess, without invoking an additional inner disk component. However, it requires and extremely round wall ($w \gg 0.4$) to achieve sufficient optical depth in the inner regions to create a near-infrared excess, which is not supported by our observations. Even though the disk wall extends over a wide radial range (cf.\ Figure~\ref{f:sbp}), the radiative transfer model requires an additional component to fit the near-infrared flux.

Since these pre-transitional disks (transitional disks with a 
near-infrared excess, implying the existence of an inner disk 
component at sub-AU radii) are thought
to be shaped by planetary systems, rather than photo-evaporation, it
is tempting to explain the the wall shape in terms of a planetary
companion. \cite{mulders13b} have done so for HD 100546 using
hydrodynamical models of planet-disk interactions, and found that the
extreme roundness ($w \approx 0.35$) in this case is best explained by
a brown dwarf companion. Taking their Figure~7 at face value, the
well-fitting range of $w=[0.19,0.36]$ for LkCa~15 spans almost the entire
plausible range of planet/star mass ratios\footnote{One would need to
take into account the difference in stellar mass between LkCa 15 and
HD 100546, which would scale down planet masses by a factor of
$\sim$2.5.}, making it difficult to \Rev{establish constraints on} planet
mass. 

We note that planet mass also affects disk morphology in other ways, 
which we have not included in our model:
\begin{description}
\item[\textbf{Gap depth.}] While our imaging data confirm an abrupt
drop in surface brightness inside the LkCa~15 disk gap, they do not 
exclude the presence of a residual level of dust in that area.  Knowing
the degree of dust depletion in the gap (the `gap depth') would impose
additional constraints on the planet mass.
However, the oversubtraction effects generated during
the ADI data reduction by the stark gradients of the nearby bright gap
render an absolute calibration of the flux level in the gap extremely
difficult.  Deep polarimetric imaging, on the other hand, may provide
access to this information in the near future.
\item[\textbf{Decoupling of gas and dust.}] In contrast to HD 100546,
the (sub)micron-sized dust grains in the disk wall of LkCa 15 show
signs of dust settling \citep{espaillat07,mulders10}, indicating that
they are dynamically decoupled from the gas and may be subject to
radial drift, as in \cite{pinilla12b}. This could alter the radial
surface brightness/density profile of the disk, and potentially
provide a better diagnostic for planet mass.

\end{description}

\subsection{\Rev{Constraints on planets and planet-induced dust grain
differentiation}}
\label{s:raddiff}

As explored in \citet{pinilla12b} and \citet{djo13}, the presence of a planet in the disk generates a radial differentiation of dust grain sizes that is strongly related to the mass of the planet. The planet causes a pressure bump to form in the gas distribution of the disk, which traps large grains but allows for small grains to filter through towards smaller radii. As a result, the distribution of large grains exhibits a wider central gap than that of the small grains.  Since NIR imaging and sub-mm interferometry mainly trace the micron- and mm-sized dust grains, respectively, this differentiation effect can be observed as a wavelength-dependent gap radius. Therefore, given the assumption that a single planet is responsible for the gap, the ratio of the gap radii measured at NIR and sub-mm wavelengths can be used to constrain the mass of the undetected planet (see Figure~8 in \citealt{djo13}). \Rev{This scenario presents a plausible explanation for the ``missing cavities'' sample of the SEEDS survey \citep{dong12} where no cavities were found in polarised intensity $H$--band images of sources displaying large cavities in sub-mm studies.}

Our $K_\mathrm{s}$-band observations, the \Rev{870}\,$\mu$m 
observations presented in \citet{andrews11}, and the SED fit carried out by \citet{espaillat10} all locate the wall at consistent radii of $r\approx[50$--$58]$\,AU. \Rev{LkCa\,15 is therefore different in this regard from the ``missing cavities'' sources.} This \Rev{lack of dust grain size differentiation} strongly suggests that the radial differentiation of dust grain sizes, if present, is not significant. Figure 8 of \citet{djo13} plots gap radius ratios between $R$-band and 880\,$\mu$m as a function of planet mass. Preliminary calculations show that replacing $R$-band imaging with $K\sub{s}$-band imaging does not significantly change this function. Taken at face value, gap radius ratios close to unity predict a mass of $M\lesssim1\,M\sub{Jup}$ for the planet responsible for the pressure distribution in the outer disc. In the case of a multi-planet system, this role is played by the outermost planet, as planets tend to influence the disk locally \citep{goodman01}.

\Rev{It may be worth to note that the gap measured in the \citet{andrews11} study, at 870\,$\mu$m ($\sim$50\,AU), is the smallest of all gaps measured at different wavelengths which is interesting since the substellar companion scenario assumed by both \citet{pinilla12b} and \citet{djo13} causes the opposite effect (i.e. larger grains are further out). In their study \citet{andrews11} do mention that their uncertainties are underestimated which could bring the value closer to the ones obtained in e.g. \citet{espaillat10} and this study, however, if confirmed, this effect would require further investigation before any conclusions regarding the cause of this ``reversed'' dust grain size separation can be drawn.}

Simulations show that a single companion would need to be very massive in order to open a gap as large as the one observed in the LkCa~15 disk \citep[$M\gg 10\,M\sub{Jup}$;][]{kraus12, crida06}.
Such a companion would be at odds with the lack of observable radial dust differentiation.
However, a system comprising multiple low-mass planets could accommodate both of these constraints.
\Citet{espaillat08} note that the inner and 
outer boundaries of the gap in the LkCa~15
disk system are comparable to the smallest and largest orbital radii in our own planetary system, 
allowing for the the tantalizing possibility that the LkCa~15 system might be a young analog of our 
Solar System.

\Citet{kraus12} propose a mass of $\sim$6\,$M\sub{Jup}$ for their planet candidate found in SAM
observations.
However, the authors note that this planet alone would be incapable of clearing the observed
disk gap, necessitating a second, lower-mass planet in a wider orbit closer to the disk gap.
Since this outer planet would dominate the radial dust differentiation, our findings do not
contradict the existence of the proposed massive planet.

We cannot directly confirm or falsify the existence of the planet 
candidate reported by \citet{kraus12} from SAM observations in $L'$
and $K$ band, since its separation from the star (70--100\,mas) lies
within the inner working angle of our ADI observations.

However, given the extreme anisotropy seen in our scattered-light images,
the possibility should perhaps not be excluded that the disk itself may
cause some of the structure seen in the SAM data, in a similar way as 
has been inferred for the similar case of T~Cha \citep{olofsson13}.

\subsection{Inner disk}
As first described in \citet{mulders10}, the nature of the inner dust component of the LkCa~15 pre-transitional disk affects the observational appearance of the outer disk, and can therefore be constrained by imaging observations of the outer disk.
An optically thick inner disk would cast a shadow on the outer disk, whereas an optically thin dust shell would not. \Rev{\Citet{espaillat11}} report a time variability in the spectrum of the LkCa~15 system in which the near- and mid-infrared fluxes exhibit negative covariance (`see-saw effect'), which is taken as evidence for time-variable shadowing, and therefore supports the optically thick inner disk scenario.

The effective scale height of the inner disk $s$ is a free parameter in our model, thus our analysis is capable of detecting variable shadowing between epochs.
Surprisingly, $\chi^2$ minimization of this parameter (cf.\ Section~\ref{s:s}) consistently yields $s=0$ for all epochs, i.e.,\ no shadowing is detected. 
This behavior can be explained either with the optically thin dust halo scenario \citep{mulders10}, with an optically thick inner disk tilted with respect to the outer disk, causing its shadow to miss the outer disk wall at most azimuths. Both explanations are, however, at odds with the `see-saw' behavior observed in the mid-infrared, which implies a photometrically significant amount of shadowing on the outer disk. Although $s=0$ is a firm result of our analysis in all epochs, it is difficult to assess why the absence of a shadow increases the fit quality, and which model parameters could alleviate the need for a fully illuminated outer disk.

\subsection{Remaining uncertainties}

We caution that our forward-modeling analysis is limited by the extent
of the disk model and its sampled parameter space.  While we consider
our best-fit models a convincing fit for our data, we acknowledge the
possiblity that they may be approximations of a disk architecture
beyond the scope of our model.

Next-generation adaptive-optics facilities may offer the means to 
reveal the structure of the LkCa~15 disk gap with more conservative
high-contrast imaging methods, resolving these remaining uncertainties.
Polarimetry is a particularly promising technique for such applications,
as demonstrated by \citet{quanz13}.


\section{Conclusion}

Following up on our $H$-band results from \citetalias{paper1} (epoch H1),
we have obtained four new epochs of $K\sub{s}$-band high-contrast 
direct-imaging observations
of the LkCa~15 system in order to reveal new insights on the 
architecture of its pre-transitional disk.  
All four $K\sub{s}$-band datasets (epochs K1--K4) were taken in 
consistent observing modes so as to
enable direct comparison, using Gemini NIRI with pupil tracking to 
enable ADI data reduction.

In order to compensate for the known flux loss and morphological 
impact of ADI, we have performed an extensive forward-modeling 
analysis.  By simulating a parametric grid of scattered-light images 
of disk models, forward-modeling them through the ADI
process, and comparing the results to the observed ADI images 
with a $\chi^2$ metric, we have arrived at stringent quantitative 
constraints on the disk geometry.  

These results are independent from, but consistent with, the 
constraints previously inferred from the system's
SED \citep{espaillat10} and from (sub-)millimeter interferometry
\citep{andrews11}.  However, they also
include new findings that have only become
accessible through near-infrared high-contrast imaging:

\begin{itemize}
\item We find
that the bright crescent seen in scattered-light ADI observations of 
LkCa~15 represents the strongly forward-scattering near side of the
outer disk, rather than the far side (cf.\ Section~\ref{s:xo}).

\item We tentatively confirm the existence of an asymmetry 
between the ansae as postulated in \citetalias{paper1}.  Under the 
assumption that the disk can be described as azimuthally symmetric 
but physically displaced along the line of nodes (the major axis of 
the projected disk gap) by an offset $x$, we find consistently 
positive best-fit values for $x$ in all four $K\sub{s}$-band epochs,
in line with our H1 results (cf.\ Section~\ref{s:xo}).
However, while the size of the best-fit $x$
offset is consistent with the H1 value in our lower-quality epochs
K2 and K4, the higher-quality data of K1 and K3 indicate a much 
smaller offset consistent with zero.  This may suggest that the 
observed asymmetry may be caused by a different mechanism than a
displaced disk gap, such as perhaps by a spiral density wave beyond 
the western ansa of the disk gap, which may be mistaken for the disk
gap in lower-quality data.

\item Furthermore, our best-fit models exhibit a surprisingly large
disk offset $y$ perpendicular to the line of nodes (along the minor
axis of the projected disk gap), bringing the brightly illuminated
near-side rim of the outer disk closer to the star (cf.\ 
Section~\ref{s:y}).  Unlike $x$, the
values of $y$ are consistent with each other and significantly 
inconsistent with zero in all four $K\sub{s}$-band epochs, lending 
the result credibility.  While such a large displacement ($\approx$0.3
times the gap radius, deprojected) constitutes a dramatic departure
from the symmetric architecture commonly assumed for LkCa 15, it 
could plausibly have escaped detection in SED and submillimeter 
interferometry.  However, we note that a displaced circular gap is no
longer a good approximation of an elliptical gap at an eccentricity
of $e=0.3$, so the real shape of the LkCa~15 disk may well lie outside
of the model parameter space considered in this analysis.

\item We find evidence for a `round' gap wall; i.e.,\ a gradual rather
than abrupt gradient of surface density between the gap and the 
outer disk, consistent with the wall being sculpted by a planetary 
companion (cf.\ Section~\ref{s:w}).

\item We obtain best results for a completely absent shadow from the 
inner disk on the outer disk wall (cf.\ Section~\ref{s:s}).  
This may indicate that the inner
disk is inclined with respect to the outer disk, causing its shadow to
miss the outer disk at most azimuths.

\end{itemize}

\noindent
The best-fit disk models are visualized in Figure~\ref{f:bestfit}, and 
the numerical results of the analysis are summarized in 
Table~\ref{t:sanity}.

Finally, we acknowledge the fact that the results of our 
forward-modeling analysis are necessarily bound by the scope of the 
model and its parameter space, and that they are therefore simplistic 
approximations to the real, complex reality of the LkCa~15 disk.  
However, given the observational status quo, we consider them to 
represent the best estimate of the disk architecture to date.

In the near future, imaging polarimetry with next-generation adaptive
optics facilities such as Subaru SCExAO \citep{guyon11} or SPHERE ZIMPOL 
\citep{thalmann08} may provide
a more detailed view of the LkCa~15 disk with fewer model dependencies.


\begin{acknowledgements}
We thank F.~Meru and S.~Quanz for useful discussions\Rev{, and the 
anonymous referee for helpful comments that improved the quality of the
manuscript.}
CTh is supported by the \Rev{European Commission under the} Marie Curie IEF 
grant No.~329875.
CAG is supported by the U.S.\ National Science Foundation under Award 
No.~1008440 and through the NASA Origins of Solar 
Systems program on NNG13PB64P. 
MM acknowledges funding from the EU FP7-2011 under Grant Agreement No.~284405.
JC is supported by the U.S.\ National Science Foundation under Award No.~1009203.
\Rev{This work is based on observations obtained at the Gemini Observatory, which 
is operated by the 
Association of Universities for Research in Astronomy, Inc., under a cooperative agreement 
with the NSF on behalf of the Gemini partnership: the National Science Foundation 
(United States), the National Research Council (Canada), CONICYT (Chile), the Australian 
Research Council (Australia), Minist\'{e}rio da Ci\^{e}ncia, Tecnologia e Inova\c{c}\~{a}o 
(Brazil) and Ministerio de Ciencia, Tecnolog\'{i}a e Innovaci\'{o}n Productiva (Argentina).
The authors wish to recognize and acknowledge the very significant cultural role and reverence that the summit of Mauna Kea has always had within the indigenous Hawaiian community.  We are most fortunate to have the opportunity to conduct observations from this mountain.}
\end{acknowledgements}



\clearpage

\begin{appendix}


\raggedbottom

\section{Preliminary best-fit solutions}
\label{a:pbf}

The parameter sets listed in Table~\ref{t:pbf} represent the preliminary,
coarse model
fits to our four observing epochs of LkCa~15 obtained through an iterative
search during the early phase of this work.  They are used as a starting
point for the studies presented in Sections~\ref{s:xo} and \ref{s:s}.
Our final, optimized solutions and their confidence intervals 
are described in Section~\ref{s:bestfit}.

\begin{table}[h!]
\caption{Description of the preliminary best-fit (PBF) solutions to the 
	four datasets.}
\label{t:pbf}
\begin{tabular}{@{}lrrrr@{}}
Parameter &  K1	& K2   & K3 & K4 \\[1mm]
\hline
Scale factor $f$ 			& 0.93	& 1.00	& 1.00	& 0.97\\
Gap radius $r$ (AU) 			& 52		& 56		& 56		& 54\\
Inclination $i$ ($^\circ$)	& 52		& 50		& 48		& 48\\
Henyey-Greenstein number $g$		& 0.60	& 0.80	& 0.70	& 0.60\\
Settling parameter $s$		& 0		& 0		& 0		& 0\\
Wall roundness $w$			& 0.25	& 0.35	& 0.25	& 0.30\\
Orientation angle $o$		& 147.7	& 150.5	& 150.7 	& 150.7\\
Center offset $x$ (mas)		& 12		& 73		& 10 	& 45\\
Center offset $y$ (mas)		& $-68$	& $-88$	& $-84$	& $-88$\\
\hline
\end{tabular}
\end{table}


\section{Best-fit solutions under the restriction of $\boldsymbol{y=0}$}
\label{a:yzero}

Our analysis clearly favors a significant offset $y$ of the disk
center perpendicular to the line of nodes (cf.\ Section~\ref{s:y}).
However, this offset is as of yet unconfirmed, and could therefore 
represent an unknown bias in our modeling approach.  For this reason, 
we here provide an overview of how the imposed restriction of $y=0$ 
would affect our results.

\begin{figure}[ptb]
\centerline{\includegraphics[width=\linewidth]{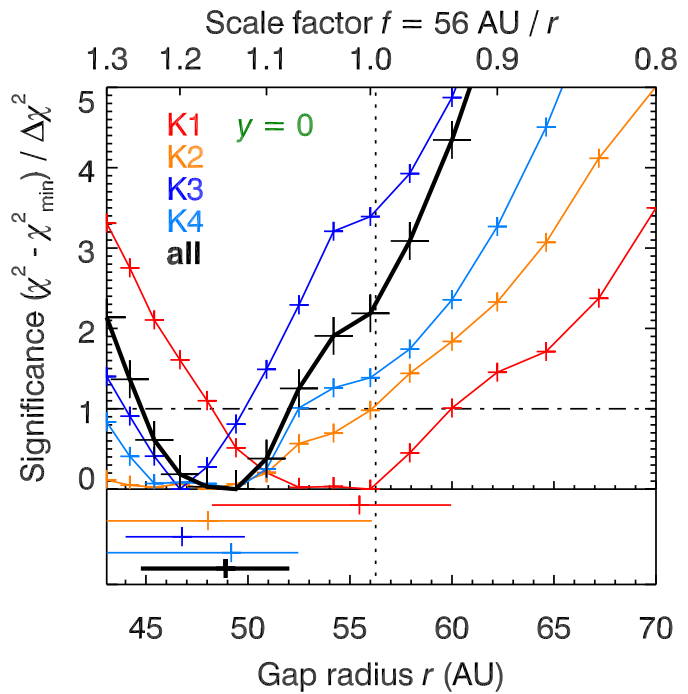}}
\caption{Constraints on the outer disk wall radius $r$ under the 
	restriction of $y=0$.
	The plot shows the excess of the best-fit $\chi^2$ for a given $r$ 
	with respect to the global minimum $\chi^2_\mathrm{min}$, normalized
	by the threshold value $\Delta\chi^2$.  The color-coded curves 
	represent each individual epoch, whereas the thick black curve is
	obtained by combining all four epochs. 
	The dotted vertical line marks the best-fit value of $r$ from the
	unrestricted analysis.
	\Rev{The bottom panel shows the well-fitting range and the best-fit value
	for all five analyses.}
	}
\label{f:r_yzero}
\end{figure}

\begin{figure}[ptb]
\centerline{\includegraphics[width=\linewidth]{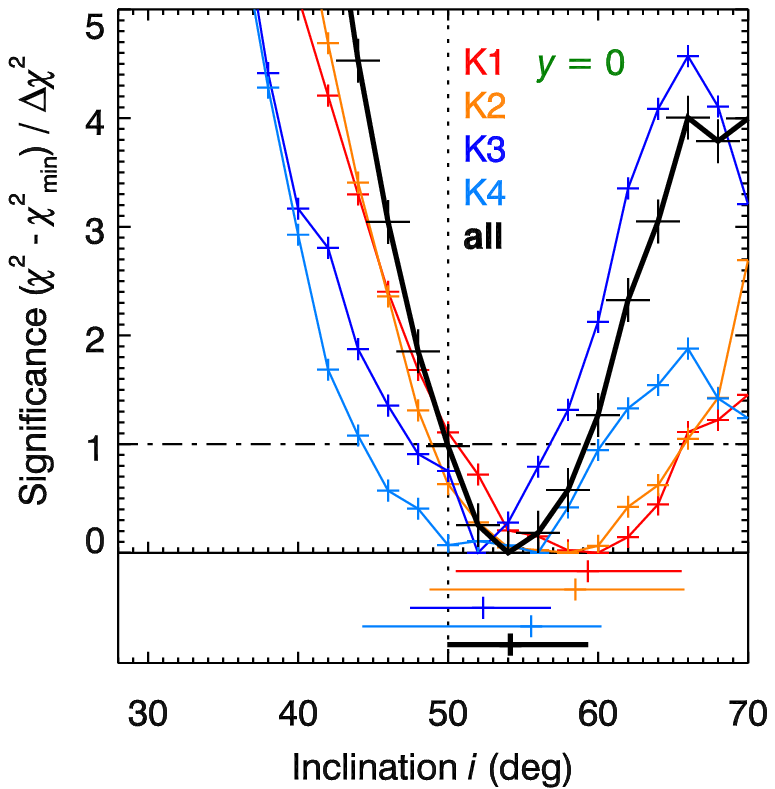}}
\caption{Constraints on the inclination $i$ of the disk plane under
	the restriction of $y=0$.
	The plot shows the excess of the best-fit $\chi^2$ for a given $i$ 
	with respect to the global minimum $\chi^2_\mathrm{min}$, normalized
	by the threshold value $\Delta\chi^2$.  The color-coded curves 
	represent each individual epoch, whereas the thick black curve is
	obtained by combining all four epochs. 
	The dotted vertical line marks the best-fit value of $i$ from the
	unrestricted analysis.
	\Rev{The bottom panel shows the well-fitting range and the best-fit value
	for all five analyses.}
	}
\label{f:i_yzero}
\end{figure}

\begin{figure}[ptb]
\centerline{\includegraphics[width=\linewidth]{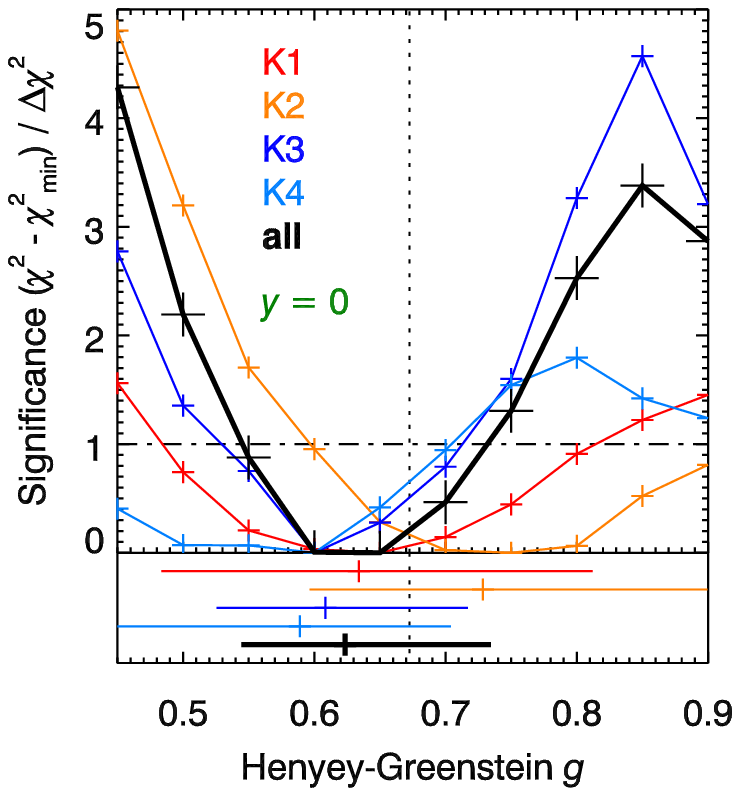}}
\caption{Constraints on the Henyey-Greenstein parameter $g$ under
	the restriction of $y=0$.
	The plot shows the excess of the best-fit $\chi^2$ for a given $g$ 
	with respect to the global minimum $\chi^2_\mathrm{min}$, normalized
	by the threshold value $\Delta\chi^2$.  The color-coded curves 
	represent each individual epoch, whereas the thick black curve is
	obtained by combining all four epochs. 
	The dotted vertical line marks the best-fit value of $g$ from the
	unrestricted analysis.
	\Rev{The bottom panel shows the well-fitting range and the best-fit value
	for all five analyses.}
	}
\label{f:g_yzero}
\end{figure}

\begin{figure}[ptb]
\centerline{\includegraphics[width=\linewidth]{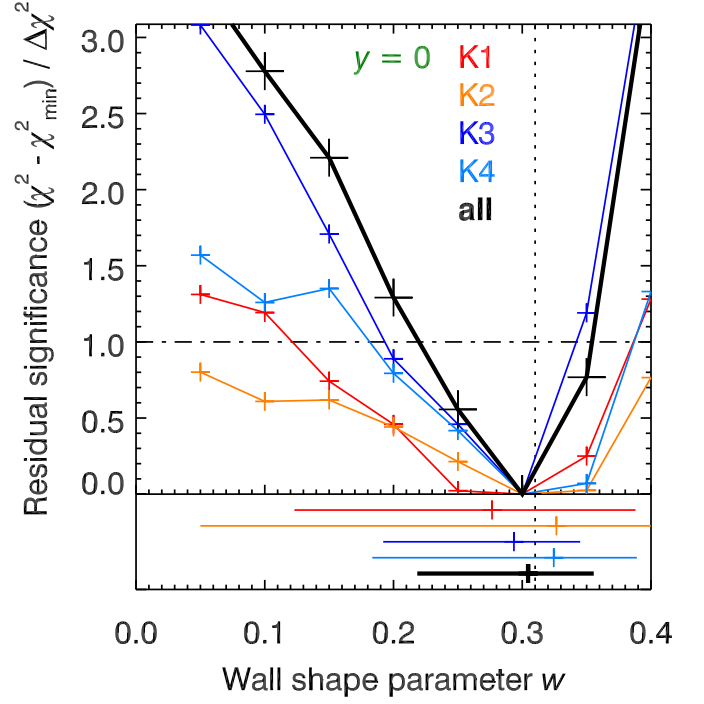}}
\caption{Constraints on wall shape parameter $w$ under
	the restriction of $y=0$.
	The plot shows the excess of the best-fit $\chi^2$ for a given $w$ 
	with respect to the global minimum $\chi^2_\mathrm{min}$, normalized
	by the threshold value $\Delta\chi^2$.  The color-coded curves 
	represent each individual epoch, whereas the thick black curve is
	obtained by combining all four epochs. 
	The dotted vertical line marks the best-fit value of $w$ from the
	unrestricted analysis, which coincide with the values obtained 
	under $y=0$.
	\Rev{The bottom panel shows the well-fitting range and the best-fit value
	for all five analyses.}
	}
\label{f:w_yzero}
\end{figure}

Since some covariances are found between the parameters $r, i, g, w$ 
and $y$ (cf.\ Section~\ref{s:covar}), the 
$\chi^2$ plots for those parameters undergo significant changes.  
Figures~\ref{f:r_yzero}--\ref{f:w_yzero} show the corresponding plots
for the $y=0$ parameter sub-space. 

In the unrestricted analysis, a range of $y$ values are included in
the well-fitting family, each of which contributes a narrow valley to
the $\chi^2$ landscape.  For parameters that covary with $y$, the
$\chi^2$ valleys change their positions with varying $y$, thus the
final $\chi^2$ valley calculated over all $y$ is broadened.  For this
reason, imposing the restriction of $y=0$ results in steeper $\chi^2$ 
curves than in the unrestricted case.

Furthermore, since the natural best-fit value of $y$ is negative, the
restriction of $y=0$ causes a negative [positive] shift in the 
best-fit value of parameters that exhibit positive [negative]
covariation with $y$.  The gap radius $r$ decreases from 56\,AU to
50\,AU, whereas the inclination $i$ rises from $50^\circ$ to 
$54^\circ$.

\section{Exploration of disk wall scale height $\boldsymbol{h}$ as an
additional free parameter}
\label{a:h}

We ran a small-scale version of the brute-force parameter grid with 
outer disk scale heights $h=[1, 2]$ in order to test whether the 
observed $y$ offset could be an aliasing effect of an inaccurate 
assumption on $h$. Since $h=1$ represents the hydrostatic equilibrium
value, the test value of $h=2$ is an extreme case intended to probe
the possible codependence of $h$ and $y$ with high sensitivity.

The analysis was run only for the best epoch, K3.  We used the 
following parametric grid listed in Table~\ref{t:h}.  
Its irregularity stems from
the fact that it was assembled in two stages after it was found that
the first stage did not fully encompass the best-fit solution.

\smallskip
\begin{table}[h!]
\caption{Parameter grid for the $h=2$ analysis.}
\label{t:h}
\vspace*{-2mm}
\begin{tabular}{@{}l@{~~}l@{~~}l}
	 $h$ &=& [1,2]\\
	 $g$ &=& [0.3, 0.5, 0.7, 0.8, 0.9]\\
	 $s$ &=& [0, 1]\\
	 $w$ &=& [0.05, 0.10, 0.15, 0.20, 0.30]\\
	 $i$ &=& [40, 42, 44, 48, 52]\,deg\\
	 $f$ &=& [0.933, 0.967, 1.000, \ldots, 1.133]\\
	 $o$ &=& preliminary best-fit values\\
	 $x$ &=& preliminary best-fit values\\
	 $y$ &=& $[-120, -110, -100, -90, -80, -60, -40, -20, 0]$\,mas
\end{tabular}
\end{table}
\smallskip

The best-fit solution was achieved for $g=0.70$, $s=0$, $w=0.05$, 
$i=44^\circ$, $f=1.067$, and $y=-90$\,mas.  The minimum $\chi^2$
achieved was 601, which is 2.8\,$\Delta\chi^2$ above the 
$\chi^2_\mathrm{min}$ of the best-fit solution with $h=1$.  While
some of the best-fit parameters for $h=2$ deviate significantly
from their $h=1$ counterparts (most notably the wall shape 
parameter $w$), we note that the best-fit value for $y$ is still
in excellent agreement with the results for $h=1$ 
($y=-70~[-115,-45]$\,mas). This implies that there is no significant
covariance between $y$ and $h$.

This behavior can be understood in conjunction with the fact that
the best-fit models do not include a shadow from the inner disk on
the wall of the outer disk ($s=0$), and thus the wall appears as
a single bright crescent rather than two distinct thin arcs.  
Increasing the outer disk scale height $h$ makes the crescent wider
(thus perhaps explaining the reduction in the wall shape parameter
$w$, which is responsible for widening the crescent in the global
best fit with $h=1$), but does not affect its overall position. 

Thus, we conclude that the observed offset in $y$ is a robust 
result of our analysis within the framework of our disk model, 
and does not reflect an underlying error
in the outer disk scale height $h$.  We continue to use $h=1$
as a fixed internal parameter for this work.

\end{appendix}

\end{document}